\documentclass[sigconf,authorversion,nonacm]{acmart}
\usepackage{bm}
\usepackage{mathtools}
\usepackage{amsmath}
\usepackage{multirow}
\usepackage{xcolor}
\usepackage{booktabs}

\usepackage{multirow}
\usepackage{makecell}
\usepackage{stfloats}

\usepackage{graphicx}
\usepackage{color}
\usepackage{textcomp}
\usepackage{cuted}
\usepackage{enumerate}
\usepackage{enumitem}
\usepackage[all]{hypcap}    
\usepackage{subfigure}
\usepackage[utf8]{inputenc}
\usepackage{booktabs}

\AtBeginDocument{%
  \providecommand\BibTeX{{
    \normalfont B\kern-0.5em{\scshape i\kern-0.25em b}\kern-0.8em\TeX}}}

\setcopyright{none}
\acmYear{2023}
\acmConference[TAICHI '23]{The 9th Annual Conference of Taiwan Association of Computer-Human Interaction}{August 19-20,2023}{Taipei, Taiwan}

\renewcommand\footnotetextcopyrightpermission[1]{} 

\begin{document}
\newcommand{\projectName}{FeltingReel}

\title[\projectName{}: Density Varying Soft Fabrication with Reeling and Felting]{\projectName{}: Density Varying Soft Fabrication with Reeling and Felting}

\author{Ping-Yi Wang}
\affiliation{
  \institution{National Taiwan University}
  \streetaddress{No. 1, Sec. 4, Roosevelt Rd.}
  \city{Taipei}
  \country{Taiwan}
  \postcode{10617}
}
\email{ping-yi.wang@hci.csie.ntu.edu.tw}

\author{Lung-Pan Cheng}
\orcid{0000-0002-7712-8622}
\affiliation{
  \institution{National Taiwan University}
  \streetaddress{No. 1, Sec. 4, Roosevelt Rd.}
  \city{Taipei}
  \country{Taiwan}
  \postcode{10617}
}
\email{lung-pan.cheng@hci.csie.ntu.edu.tw}

\renewcommand{\shortauthors}{Wang et al.}

\begin{abstract}
\textit{\projectName{}} is a soft fabrication system that allows users to create a 3D non-woven textile with various structural strengths. 
Our system coils wool yarn onto a central reel to form a basic shape and uses actuated barbed needles to refine it. 
By controlling the coiling tension and the felting times, our system varies the density of the workpiece in a target area to achieve various structural strengths. 
Specifically, our system controls the tilt of coiling and felting using a Stewart platform around a motorized rotating reel. 
Our system also allows different basic shapes with hollow internal structures to be formed by changing the detachable reel core. 
We investigate the effects of different felting needles, frequencies, and coiling directions that influence the density, structural strength, and fabrication time of a workpiece.
We propose three methods to combine felting and reeling.
We evaluate their performances and final products by producing two example workpieces using our system. 
We demonstrate several objects made by our working system and discuss its capabilities and limitations. 

\end{abstract}

\ccsdesc[500]{Human-centered computing~Human computer interaction (HCI)}

\keywords{Fabrication; felting; coiling; soft materials; computational crafts; interactive devices; reel; textile;density variation; prototype; needle felting}

\begin{teaserfigure}
  \includegraphics[width=\textwidth]{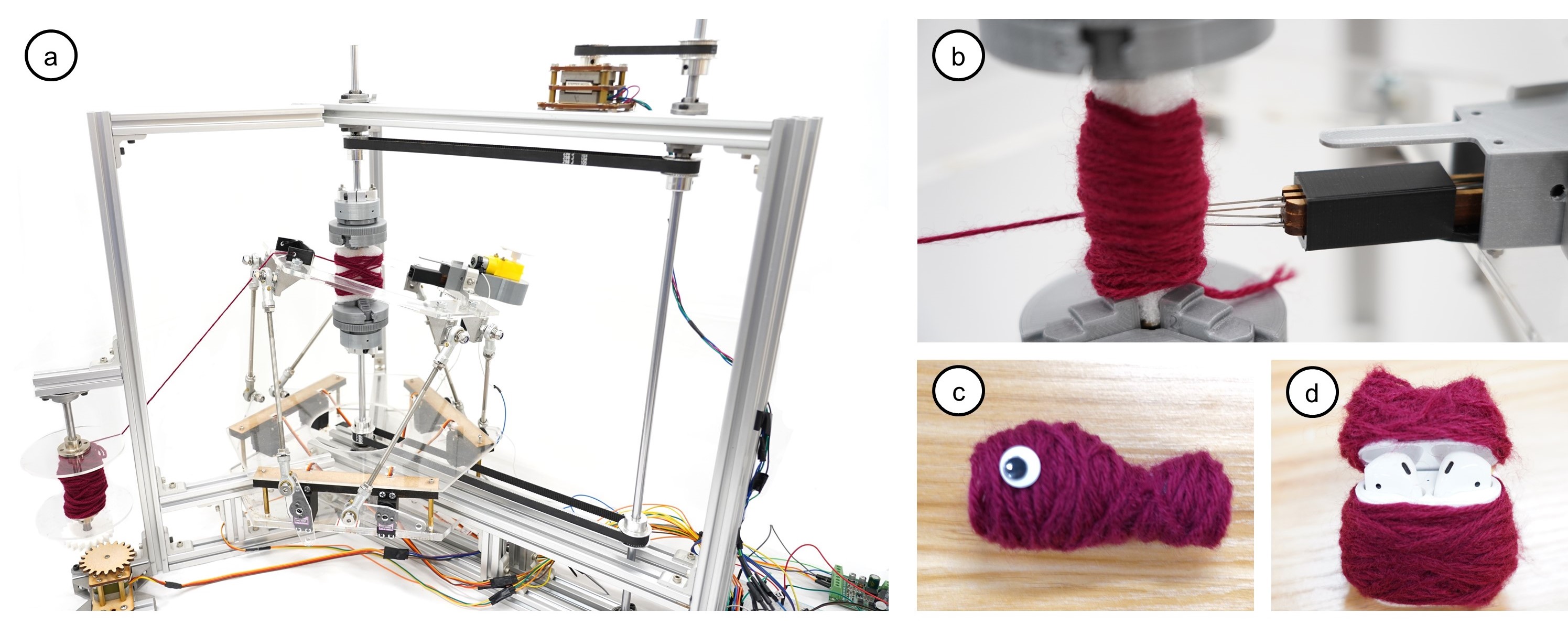}
    \caption{ \textit{\projectName{}} is a soft fabrication system that utilizes coiling and felting techniques to produce 3D non-woven textiles. (a) \textit{\projectName{}} uses the Stewart platform to position the thread feeder for coiling the textile onto the motorized reel and the felting machine for punching specific areas of the textile. (b) \textit{\projectName{}} coordinates the coiling and felting processes on the reel core to create 3D shapes and refine their surface details. Examples of objects created using this fabrication system include (c) a Fish doll and (d) an earphone case cover.}
  \Description{Figure 1 shows the teaser figure of our soft fabrication system FeltingReel with four photos labeled (a), (b), (c), and (d). Image (a) represents the whole device of FeltingReel, which uses the Stewart platform to position the thread feeder for coiling the textile onto the motorized reel in the center and enables the felting machine to target specific areas of the textile. Image (b) shows the detail of the coiling center and the felting machine by demonstrating the wool yarn coils a cylinder on the center cushion that is fixed by the two three-jaw chucks and the felting needles punch on the yarn to shape the workpiece. Examples of objects created using this fabrication system include (c) a Fish doll and (d) an earphone case cover that covers the earphone case.}
  \label{fig: teaser}
\end{teaserfigure}

\maketitle

\section{Introduction}

Textiles play an integral role in our daily lives. 
They are used to create pants, clothes, masks, bags, shoes, and even the stuffed animals we hug. 
Textiles can be broadly classified into three categories based on their manufacturing process: woven, knit, and non-woven. 
Woven fabrics are created by interlacing yarns at right angles to each other to form a structured and organized fabric (warp and weft). 
Knit fabrics are produced by interloping one or more sets of yarns, giving them stretch and flexibility.
Non-woven fabrics are made by tangling fibers together in a disorganized manner.

Woven and knit fabrics are widely used for creating clothing and household items due to their structured and organized nature. 
Many researchers focus on improving the manufacturing processes of woven and knit fabrics or combining them with other techniques such as circuits to provide additional functionalities~\cite{hofmann2020knitgist,albaugh2021engineering,albaugh2019digital,guo2020representing,kim2022knitskin,goudswaard2020fabriclick}.
However, these fabrics are often limited to 2D patterns and lack 3D volumetric shapes.
Industrial sewing and embroidery machines can create 2D fabrics with some textural protrusions but cannot produce complex 3D shapes~\cite {griepentrog2007needle}.

In contrast to woven and knit fabrics, non-woven fabric lacks organization but offers greater freedom for reforming into various 3D volumetric shapes.
This property aligns well with recent research in computational fabrication, such as 3D printing and metamaterials.
However, only a few researchers have explored how to form 3D shape materials or use non-woven fabric for fabricating their work~\cite{Review3DFabric,peng2015layered,jones2021punch}.

\begin{figure}[htb]
  \includegraphics[width=1\columnwidth]{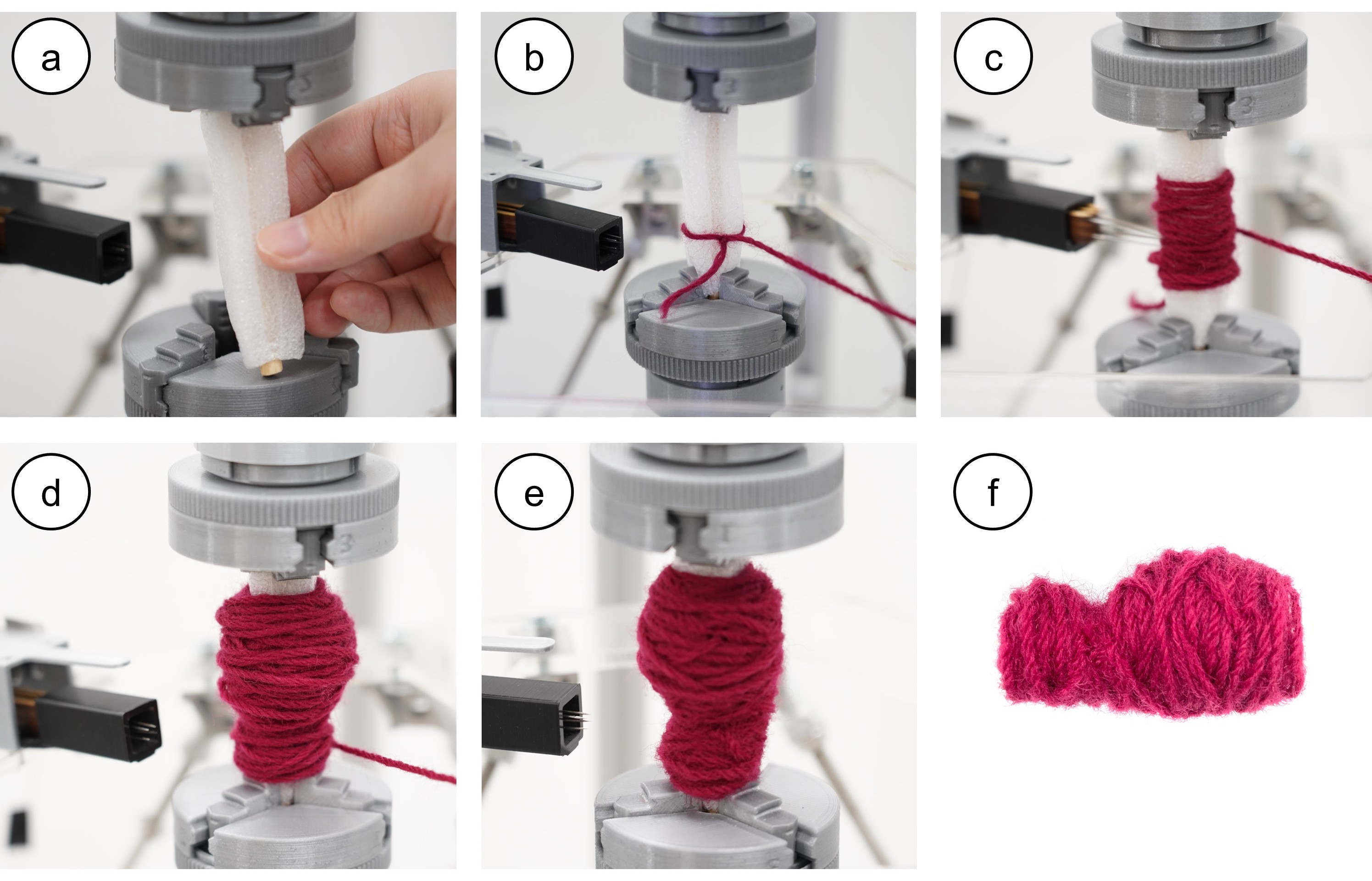}
    \caption{ The fabrication steps of making a fish doll:
    (a) install the detachable reel core to the three-jaw chucks; (b) tie the wool yarn onto the core; (c) start the system to coil and felt; (d) coil to form the basic shape; (e) felt to refine the shape. (f) The final product.}
  \Description{ Figure 2 shows the fabrication steps by FeltingReel for making a fish doll:
    (a)Install the detachable reel core to the three-jaw chucks in the center. (b) Tie the wool yarn onto the core. (c) Start the system to coil and felt. (d) Coil to form the basic shape. (e) Felt to refine the shape. (f) The final fish doll. }
  \label{fig: stream}
\end{figure}

Inspired by a lathe, we look into using felting and coiling to fabricate 3D non-woven fabric objects. 
With the non-woven property, we are able to vary the structural strength by controlling the textile density. 
We present \projectName{}, a lathe-like system that uses a reel and a felt to fabricate 3D textiles.

We use the fabrication process of a fish doll as an example to explain how \projectName{} works. 
In the beginning, we install a detachable reel core, which is made of a cushion and a stick, at the central axle(\autoref{fig: stream}a). 
We put the yarn onto the thread spool (\autoref{fig: teaser} bottom left), wire it through the thread feeder (\autoref{fig: Stewart}), and tie a knot on the reel (\autoref{fig: stream}b).
The system then starts coiling the yarn onto the reel by rotating the central axle to form a base layer(\autoref{fig:  stream}c).
The system briefly felts the workpiece to achieve fixation during the coiling process.
The system also allows users to increase the felting frequency to strengthen the structure. 
To further increase the whole object's structural strength, the system coils horizontally and tilts the platform to coil obliquely on the reel. 
After making the base layer, the system starts coiling more yarn at the bottom to form the cylindrical tail part and at the top to form the ellipsoid body part (\autoref{fig: stream}d).
The system then felts more numbers of punching to refine the curved part on one side of the fish's body and the flat belly on the other (\autoref{fig: stream}e).
Finally, we shear the wool yarn and remove the result from the reel core (\autoref{fig: stream}f).
Optionally, we can close both ends of the fish by additional manual felting.
 
Our main contribution is the soft fabrication system that creates a non-woven fabric. 
The key idea is to combine coiling and felting and use the volumetric property of non-woven fabrics to vary the density and structural strength. 
Our system provides soft fabrication more freedom to form in 3D and is able to vary the structural strength during the fabrication process with the change in coiling tension, coiling pattern, and the use of felting needles.
We demonstrate several objects made by our working system and discuss its capability and limitation.

\section{Related Work}
Our project delves into fabricating soft materials in 3D using coiling and felting techniques.
The soft fabrication field encompasses various aspects such as fabrication methods~\cite{Review3DFabric, plotting2020thread, cherenack2010woven}, adding new functionalities~\cite{Review3DFabric, kim2022knitskin}, materials~\cite{3DPrintedFabric,forman2020defextiles}, user interaction~\cite{hamdan2018sketch, del2023punchprint}, digital fabrication~\cite{ albaugh2019digital,goudswaard2020fabriclick}, etc.
Our work draws inspiration from other fabrication processes and materials and specifically focuses on exploring 3D fabrication using soft materials.
In particular, our work is primarily related to (1) soft fabrication in 3D, (2) 3D manufacturing, (3) felting, and (4) coiling.

\subsection{Soft fabrication in 3D}
While there has been significant research on soft fabrication, the majority of it focuses on manufacturing woven or knit fabrics~\cite{hofmann2020knitgist,albaugh2021engineering}, which have a more regular structure.
Alternatively, some researchers have combined other techniques, such as sensors, to add functionality to the woven fabric~\cite{albaugh2019digital,guo2020representing,luo2021knitui,kim2022knitskin}. 
There has been less discussion on the use of non-woven fabric, with most of the existing research focusing on the material properties~\cite{zoran2013freed,schofield193821,hearle197225,van1955study,van1956relation}.  

While machines that can fabricate soft materials in 3D exist in the aerospace and military industries, they are typically complex and not widely available for other applications~\cite{WeavingShoes}.
As a result, 3D soft fabrication remains primarily a research topic.
Some researchers, such as Wu et al., have explored weaving 3D objects using specific patterns for each layer~\cite{wu2020automatic}.
Others, such as Harvey et al., have used existing machines to add functional knitting~\cite{WeavingShoes}.
However, as seen in previous works~\cite{Review3DFabric,bhushan2017overview,kim2022knitskin}, most current research on 3D soft fabrication relies on large machines designed for weaving or knitting fabrics.

Various methods to fabricate non-woven fabrics include chemical, hot, and pressure techniques.
However, only a few works have utilized non-woven fabrics for 3D fabrication.
For instance, the Printing Teddy Bear project utilizes needle felting to fix yarn onto the base and stack out object shapes~\cite{hudson2014printing}, while Peng et al. stack non-woven fabric slices to form objects~\cite{peng2015layered}.

The advantage of soft fabrication is that it can give the results more freedom to bend or provide haptic feedback. 
The woven fabric and knit fabric machines have high complexity and space requirements.
Besides, the results' structural strength is determined at the beginning of the manufacturing process due to the fixed fabricate method and the fixed material.
In contrast, commercial 3D printers can change the internal structure, reduce fabrication times, and enable special functions such as rotation torque and flexibility.
One advantage of non-woven fabric is the ability to incrementally adjust the structural strength during fabrication by varying the density.
We believe non-woven fabric can be a viable solution for building a simple 3D fabrication prototype system.

\subsection{3D Manufacturing}
Most printing techniques utilize the additive manufacturing process, which involves building the structure layer by layer on a platform, much like building blocks~\cite{ngo2018additive,bhushan2017overview}.
However, these techniques are typically limited to growing structures from a flat plane and may require additional effort to handle overhanging structures~\cite{loterie2018volumetric}.
Additionally, the results cannot be refined after printing, as the modeling process determines the final outcome.
This means that if a mistake is made during modeling, the user must revise the entire model and reprint it.
Thus, it can be challenging to make revisions to the final product once it has been printed.

Few works have explored non-layer printing techniques~\cite{wang2019line,mitropoulou2020print, RoMA,shiwarski2021emergence,nonlayer5DOFwireframeprinter}.
These approaches require significant effort to calibrate the printing process and to develop algorithms that reduce fabrication time.
However, they offer users greater freedom in creating their work.
Additive manufacturing techniques can utilize a range of materials, including plastic, paper, ceramics, concrete, chocolate, and others~\cite{forman2020defextiles,miyatake2021flower,ngo2018additive,bhushan2017overview}.
Different materials result in different structural strengths, which are only changeable before the fabrication begins.

Instead of using additive manufacturing, some researchers have explored other techniques for creating 3D shapes.
One common method is traditional CNC milling.
Other techniques, such as RoboCut, utilize a flexible hot wire to carve foam and similar materials~\cite{duenser2020robocut}, while FreeD empowers creators to use a drill for sculpting and milling materials to produce objects~\cite{zoran2013freed}.
These methods enable the swift formation of fundamental shapes and allow users to modify their work, but they are unable to restore the removed material after it has been subtracted.

A few works have explored combining two different manufacturing processes to fabricate 3D objects, such as the LOM technique~\cite{park2000characterization}.
This approach involves using laser cutting to cut the contour of paper or other flat materials in each layer and then bonding the layers together with an adhesive.
Peng et al. also use laser cutting to cut the 2D contour on a fabric sheet, which is then stacked onto the previous layer using a heat-sensitive adhesive to form the 3D shape material in each layer~\cite{peng2015layered}.
In both cases, the material is subtracted first and then added layer by layer.
While there are many ways to print 3D objects, refining the shape or structural strength after printing can be challenging.
As a potential solution, we are exploring the felting technique.

\subsection{Felting}
Felting is a traditional fabrication technique for creating non-woven fabrics that have been used for centuries in both industry and handicrafts~\cite{griepentrog2007needle,chen2016three}.
Using just needles, users can punch in and out the fibers to fixate them by entanglement and refine the 3D shape by compressing them.
In the textile industry, a similar process called needle punching uses a plate with needles to compress the fiber together to form non-woven textiles~\cite{jones2021punch,chen2016three}. 
McGee et al. show the needle punching examples of the soft robotic field in~\cite{mcgee2019hard+}.
In handicrafts, the process is called needle felting or wool felting, where the barbed needle is used to tangle the fibers together to shape the craft by passing it through wool roving.
Through repeated stabbings of the yarn, the fibers become more tightly intertwined, fixated, and compressed, allowing for precise shaping and deformation of specific parts of the final product.
These two properties inspire us to design our system.
While some works, such as Printing Teddy Bear, use needle felting to fixate the yarn~\cite{hudson2014printing}, our approach is different in that felting is not only used to bond the fibers together but is also used to refine the shape of the 3D object.

\subsection{Coiling}
The research about coiling discusses the mechanism for winding the motor's coil~\cite{toliyat1991analysis} or relevant physical properties~\cite{mertiny2002influence}.
Most shape is simple and just coil the thread together~\cite{5DOF2016CoilingPrinter}. 
Besides forming the particular shape for the motor's coil or collecting thread on a spool, researchers seldom discuss whether to fabricate 3D by coiling.
However, the manufacturing process of coiling is rapid.
In our work, we give coiling as another fabrication method for forming the soft material to find its capability and limitation.

\subsection{Summary}
Currently, few soft fabrication works can change the structural strength during or after the fabrication process.
We see the possibility of using the needle felting technique and varying density coiling of soft fabrication to vary the structural strength during the fabrication.
Furthermore, few soft fabrication works have the ability to form 3D models.
The coiling method will allow us to form the base of a 3D workpiece rapidly, and the felting will enable us to finely shape the object.
With these two methods, we can provide soft fabrication with more freedom.
We look into how the system can fabricate soft material by coiling and felting.

\section{Design Rationales}
Our primary goal is to build a device to fabricate 3D soft objects by coiling and felting non-woven fabrics. 
Coiling is used to form the basic shape of the soft material with a controllable source.
Felting is used to fixate and reform the basic shape from its surface.
We describe our design rationales in detail as follows. 

We use a Stewart platform to control the coiling and the felting direction. 
Unlike traditional 3D printers that use an x-y table to form the 2.5D object~\cite{ngo2018additive,mackay2017performance,yan2021fabhydro}, a Steward platform has more degrees of freedom in moving and orienting.
We need at least 5 DoF to position the felting needles at an angle away from perpendicular to the workpiece surface at arbitrary points. 

We discard using the robotic arm due to the high controlling effort and the platform's stability.
Robotic arms tend to have very long torque arms. 
When applying force at the end effector, here, the felting machine, the felting position would not be stable unless using the powerful robotic arm that tends to raise their cost and size. 
The felting machine requires stable support on all freedoms except the freedom of moving needles.
The slight movement of the needle moving direction is tolerable in our design.


A lathe is a machine that contains a rotating central axle to place the object and a chisel cut around the axle. 
This structure inspires the design of our device.
We place a coiling axle in the center of the platform to fabricate our workpiece on it. 
The coiling platform controls the rotation and provides enough torque to coil the fabric thread.
When coiling, the platform requires a mechanism to feed the fabric thread at the right pitch. 
The mechanism provides a suitable quantity of fabric thread and maintains the tension of the line at the same time.
Suitable fabric thread that can be felt together also has to be considered.

Besides, the working space (reel core) on the rotating central axle must be detachable so we can remove the workpiece from the system. 
The additional benefit of this design is the capability to change the core's shape.
This design can expand the shape that the system can achieve.
For example, we can form a shoe by changing the core shape to a shoe support. 
To provide a detachable working space, we also need the structure to fixate it when working and open the tool to remove the work.

To facilitate the repetitive felting process, we need a mechanism to control the needle punching where (1) we can stably install the needles on it and replace them from the device because those barbed needles are consumables, (2) it is small and light enough to place on the motion platform, (3) it can let the needles move back and forth repeatably and rapidly and (4) it provides enough force to stab through the thread. 

Finally, we aim for a working prototype without optimizing its efficiency and accuracy, so we choose a standard microcontroller to connect our software and hardware components. 

To sum up, we choose (1) a motion platform that can give enough freedom to control the system, (2) a motorized rotating axle with a replaceable core and motorized reels that can adjust coiling tension for stable material supply, (3) a felting machine that can automatically and rapidly felt material, and (4) a basic control system to connect our software and hardware.

\section{System Implementation}
\begin{figure}[htb]
\includegraphics[width=1\columnwidth]{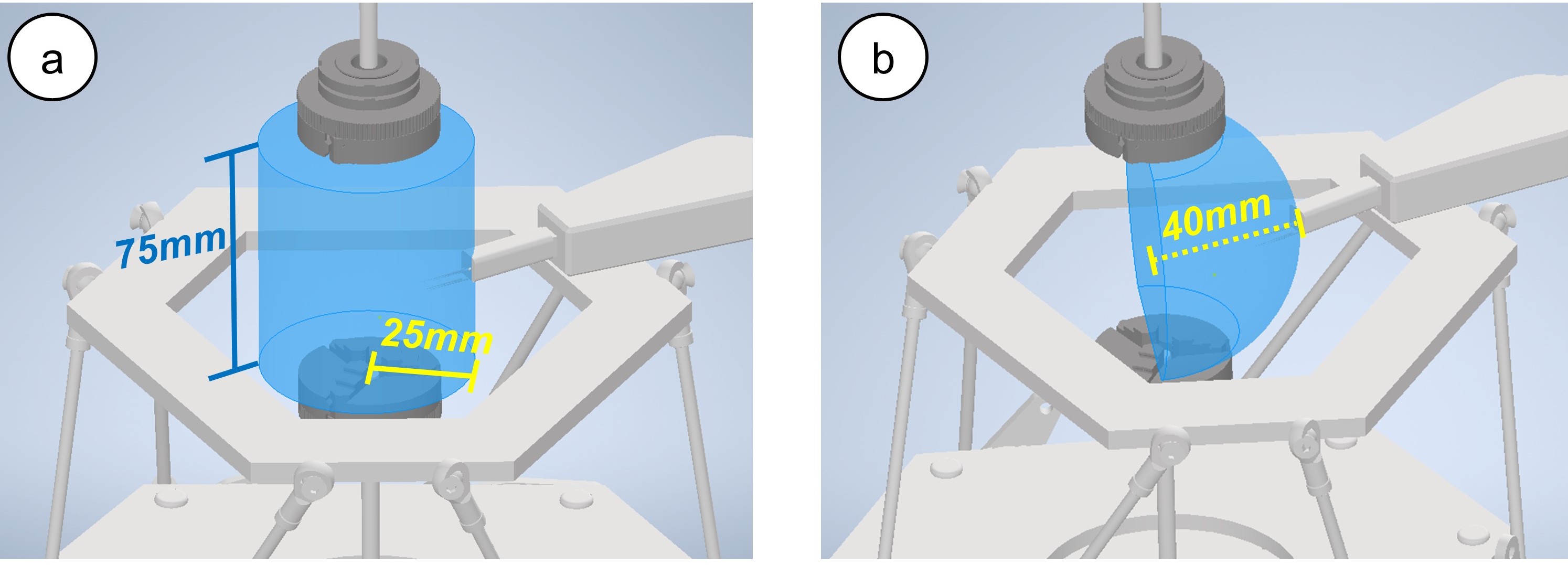}
    \caption{(a) The blue cylinder area represents the working area (radius: 25 mm, height: 75 mm) in our system. (b) The blue hemisphere represents the operating range of the felting machine. (radius: $\approx$40 mm) }
  \Description{ Figure 3 shows the working space area by the two images labeled with (a) and (b). (a) The blue cylinder area in the center of the Stewart Platform represents the working area (radius: 25 mm, height: 75 mm) in our system. (b) The blue hemisphere in the center of the Stewart Platform represents the operating range of the felting machine. (radius: 40 mm)   }
  \label{fig: working space}
\end{figure}

\begin{figure}[htb]
  \includegraphics[width=1\columnwidth]{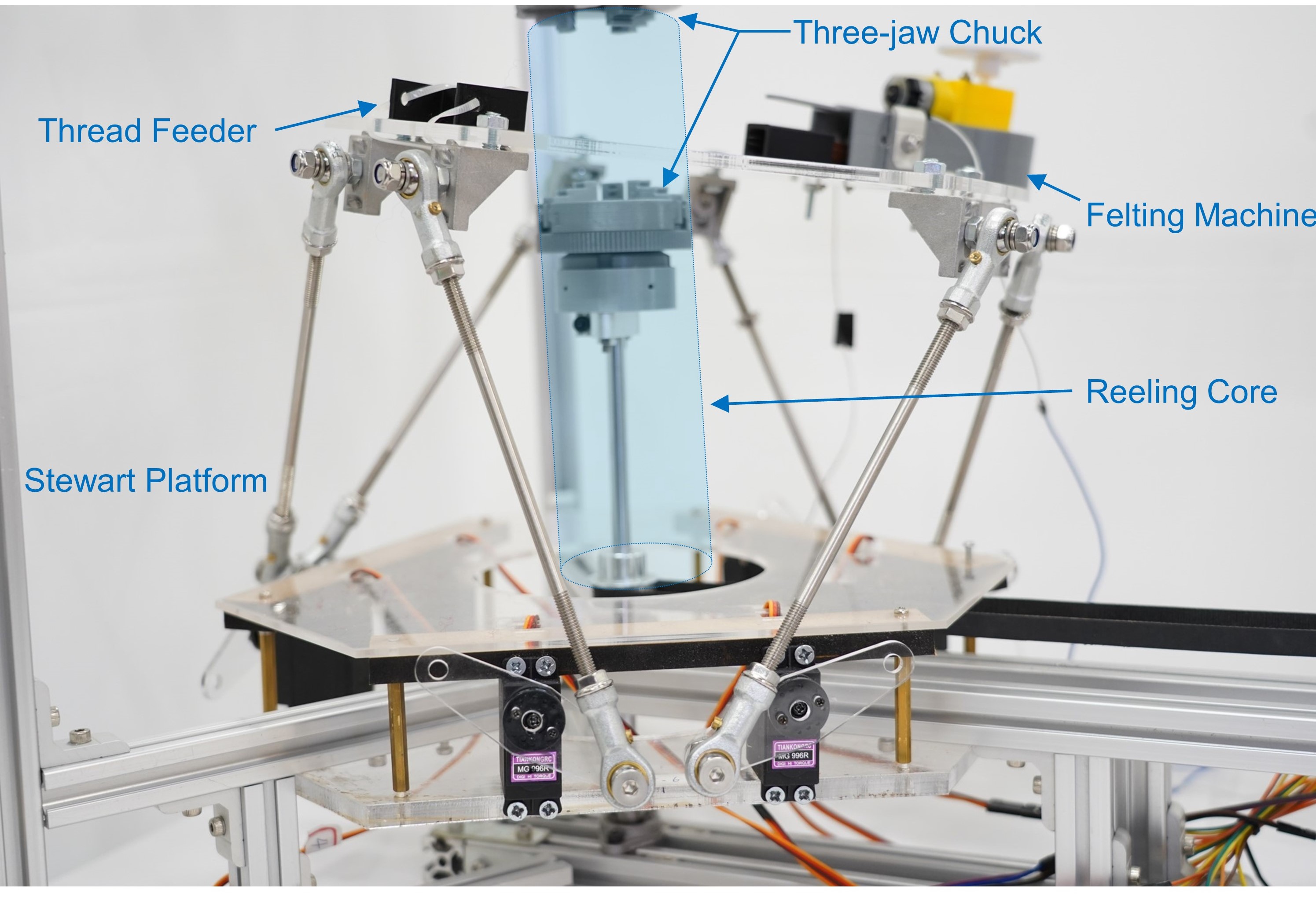}
    \caption{The Stewart platform gives our device more freedom to move our felting machine and thread feeder. We utilize three-jaw chucks to secure the object firmly onto the rotating axle.}
  \Description{Figure 4 shows the Stewart platform, which gives our device more freedom to move our felting machine. It is fixed on the aluminum extrusion to become more stable. The figure labeled the position of the rotation center in the center and the two three-jaw chucks. Also, the image shows that the felting machine and feeder were placed on the two sides of the platform. }
  \label{fig: Stewart}
\end{figure}

We build the hardware system using the aluminum extrusion as the supporting frame to assemble each component (\autoref{fig: teaser}a and \autoref{fig: Stewart}). 
It consists of (1) a motion platform, (2) a felting machine, (3) a motorized reel axle, (4) motorized reels for tension control, and (5) a control board. 
All components are mounted onto the aluminum frame to strengthen the stability. 
The maximum working range is a cylinder with a diameter of 6 cm and a height of 10 cm (\autoref{fig: working space}a). 
Currently, the shape made by our system is limited by the size of the working space.
All the shapes start as a convex hull winding around a core with the limited size shown in \autoref{fig: working space}a.
After winding, the shape is modified by compressing the particular areas. 
The refined shape is limited by the direction and position where
 the felting needles can achieve\autoref{fig: working space}b. 

\subsection{Motion Platform}
According to our design rationales, we implement a Stewart platform using the ball joint linkage design~\cite{dasgupta2000stewart}. 
It consists of 6 servo motors (MG996R) and 6 ball joint linkages (22 cm) to achieve 6DoF control of a hollow acrylic platform (8 mm). 
The hollow acrylic platform allows tools to be mounted on the platform's rim while giving rotating space for the central axle. 
All servo motors are connected to our microcontroller. 

The platform forms a specific position and rotation according to kinematics when we set each servo with a certain angle.  
We use inverse kinematics to calculate the linkage's tilt to map the position and the orientation to the servo motor angle~\cite{dasgupta2000stewart}. 
Because each structure's dimension is constant, we simplify the operations to a transform matrix.
The Stewart platform's moving range and the felting machine's working range in \autoref{fig: working space}b also influence the max work size (\autoref{fig: working space}a).

\subsection{Reel Core and Feeding Mechanism}
We use the stepper motors (Nema 11) to control the reel core rotation and the thread tension control. 
They provide stable turns at a specific clockwise or counterclockwise angle and control the tension of the thread by maintaining the two motors rotating speed ratio.  
It also provides enough torque (0.2 N-cm) and is small enough to be placed on aluminum extrusion. 
We control the rotation at a certain angle by counting the steps (200 steps for a cycle). 
For motor drivers, we use L298N.
We use the same motors and drivers to reduce the controlling complexity.
To fix the vertical rotation axle, we use four bearings at the central axle, four at the rotating axle next to the stepper motor, and two at the thread spool.

The stepper motor is set on the sides using the belt gear driving the reel core to avoid hindering the working space. 
The feeding mechanism is placed on the side of the structure to replace the spool easier. 

Yarn made from wool is the most suitable material due to the micro-structure of its fiber surfaces.
Our system uses it as our primary material, and we also test other common thread materials such as cotton, hemp rope, or acrylic (discussed in the Evaluation section). 

The workpiece must be able to remove from the device, so the reel core is designed to be detachable. 
We use the three-jaw chuck to fix the reel core. 
Upon completing the workpiece, users have the option to remove the reel core, resulting in a hollow object, or to leave the core intact to form a solid object, depending on their desired outcome. 

There are requirements for the material property of the reel core used in the system.
When felting, the machine needs a soft material (a polyurethane foam cushion in our project) to let the barb of the needle stab in the inner fiber to form firm entanglement.
In addition, the needles will break if they keep punching the solid object. 
However, if the reel core is too soft to resist the punching force, it will bend when the needle punches it.
The material of the reel can not be too hard or too soft, so we combine a hard stick in the center of a soft cushion. 
With the support of the stick, the core will not be deformed when felting; 
with the surrounding cushion, the needles can stab into it to let the barb hook the fiber.

\subsection{Felting Machine}
We modify the off-the-shelf felting machine to meet our design requirements. 
The felting machine pushes needles back and forth using a motorized crank. 
We replace its motor and tested it with the stepper, servo, and DC motor to control the felting machine. 
While the servo and stepper motor we find can do more precise extrusion at the right time, we choose to use a DC motor to meet our torque requirement~\cite{hearle197225}
 and fit in the space on the platform. 
More powerful and smaller steppers or servo motors can be better alternatives. 
We set the felting machine on the Stewart platform and place the rotating reel core in its hollowed center. 



\subsection{Controlling System}
We provide basic functions and advanced functions that include more technical details for users to use in our system. 
The basic functions include controlling the position of the Stewart platform, the rotation cycle and speed of the stepper motors for the central axle and the thread spool, the frequency and execution time of the auto-felting machine, and the operating test of each device. 
It also includes functions about the setting of the system like (1) the line width, which influences the coiling cycles and the spacing between each cycle, and (2) the coiling tension, which is adjusted by calculating the wheel ratio with the walking steps of two stepper motors and controlling the motors moving synchronously with the corresponding speed.

The advanced functions are felting around the object, felting vertically of a certain side of the object, coiling layers horizontally, and coiling layers crossly. 
Felting a whole circle around the central axle controls the central axle to rotate and the felting machine with a given felting time and position; 
felting vertically controls the motion of the Stewart Platform and the operating time of the felting machine; 
and coiling layers horizontally or crossly lets the user input parameter of coiling density, layer number, and felting frequency. 
The function controls the execution together with all the components.
Our system also includes functions that support the 3D shapes we have made, such as balls, cylinders, phones, and fish in a certain size. 
In each function, our system sends accurate pulses to the controllers to control the angle of servo motors from the Stewart platform, the steps of stepper motors from the rotating platform, and the time of the DC motor from the felting machine.

\section{From 3D Model to FeltingReel}
\label{sec:3D}
We provide three methods to transform a user-designed model into 3D non-woven textile by different portions of coiling and felting. 
While the goal is the same---to shape those wool yarns to fit the target model, three methods take different fabrication times and result in different densities.

\begin{figure}[htb]
\includegraphics[width=1\columnwidth]{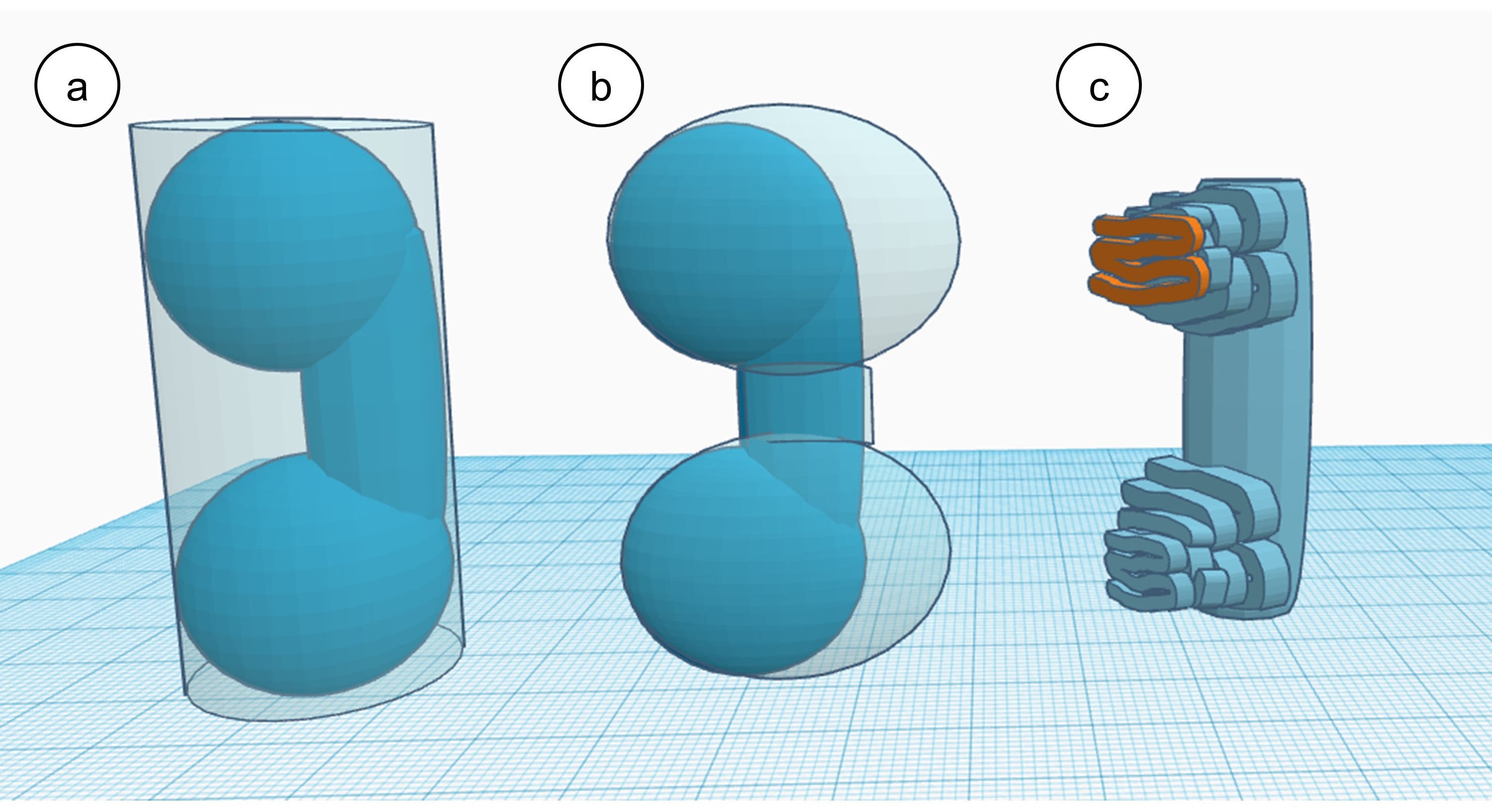}
    \caption{ The schematic of fabricating phones by the three different transform methods: (a)regular coiling, (b)fit coiling, and (c)embroidery. We use these three methods to show how we do the transformation from the model to the system. The transparent area represents the original area with coiling. The solid part shows the final result after felting. 
    The system coils a base with a cylinder shape in the (a) regular coiling method and a dumbbell shape in the (b) fit coiling method. 
    In (c) embroidery, the protruded part (especially highlighted in orange) represents the line winding and fixed on a certain side rather than coiling around.}
  \Description{ Figure 5 shows the schematic of fabricating phones by the three different transform methods: (a)regular coiling, (b)fit coiling, and (c)embroidery. (a) is on the left, showing a solid phone shape that is included in a transparent cylinder. (b) is in the center showing a solid phone shape that is included in a transparent symmetric rotation model(like a dumbbell). (c) is on the right, shows a column attached by the winding lines on the left up and left down of the column, which prototype like a phone. }
  \label{fig: three methods}
\end{figure}

\subsection{Regular Coiling Method}

The first method is called \emph{regular coiling method}, whose schematic is shown as \autoref{fig: three methods}a.  
This method forms a regular cylinder (the transparent part in \autoref{fig: three methods}a) whose size bounds the target model as the base shape and then uses felting to compress the rest part to fit the model. 
This is the easiest conversion as there is no additional process to generate code for making the bounding cylinder. 
However, the felting time is long if the model has lots of parts that need to be compressed, especially containing a deep concave shape. 

\subsection{Fit Coiling Method}

The second method is called \emph{fit coiling method}, shown in \autoref{fig: three methods}b. 
The inspiration is from the normal coiling (one-direction coiling), which forms the symmetric rotation work. 
The one-direction coiling way is like a spinning machine. 
In this method, we form the symmetric rotation model (the transparent dumbbell-shape part in \autoref{fig: three methods}b) of the target model by coiling first, then using the felting to compress the part that we do not need to fit the target model.  
The system varies the number of coiling cycle layers to form the different thicknesses corresponding to the central axle. 
The coiling mixes the crossed and horizontal coiling methods to strengthen the structure. 


\subsection{Embroidered Method}

The last method is like adding the extruded part to the base called \emph{embroidered method}, shown in \autoref{fig: three methods}c. 
This method forms a base cylinder first, while the difference to the regular coiling method is that its size is smaller than the target model. 
The reason to use the cylinder as the base shape is that it is the simplest model that can be rapidly formed by our system.
Also, the density of the cylinder can be adjusted by controlling the coiling way and coiling strength. 
By continuously changing the rotation direction and using needles to fix the yarn on the base, the system embroiders the yarn onto the base. 
Felting fixates the yarn, and the yarn quickly stacks back and forth in a focus region. 
This method can use a stack to form the protruding part.

\subsection{Examples}

\begin{figure}[htb]
  \includegraphics[width=1\columnwidth]{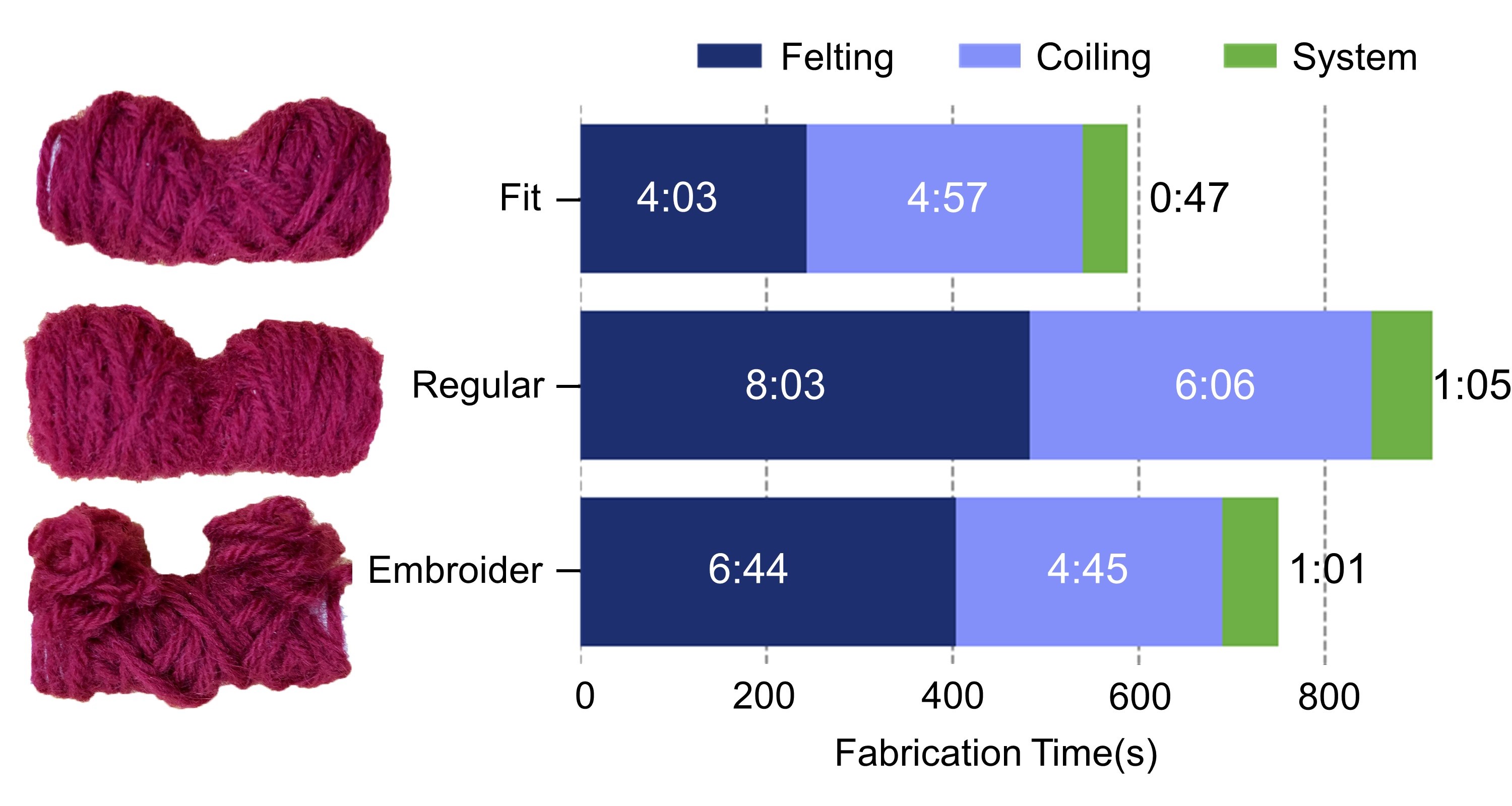}
    \caption{ We compared the fabrication times of phone dolls made using three different methods: fit coiling, regular coiling, and embroidered. The regular coiling method requires more time for felting, while the fit coiling method takes less time for total fabrication. }
  \Description{Figure 6 shows the bar plot of the fabrication times of phone cases made using three different methods: fit coiling, regular coiling, and embroidered. The right of the bar demonstrates the three workpieces made by these methods. The felting time is 4:03 (4m3s), 8:03, and 6:44, respectively; the coiling time is 4:57, 6:06, and 4:45; the system time is 0:47,1:05 and 1:01.  }
  \label{fig: phone visualization}
\end{figure}

\begin{figure}[htb]
  \includegraphics[width=1\columnwidth]{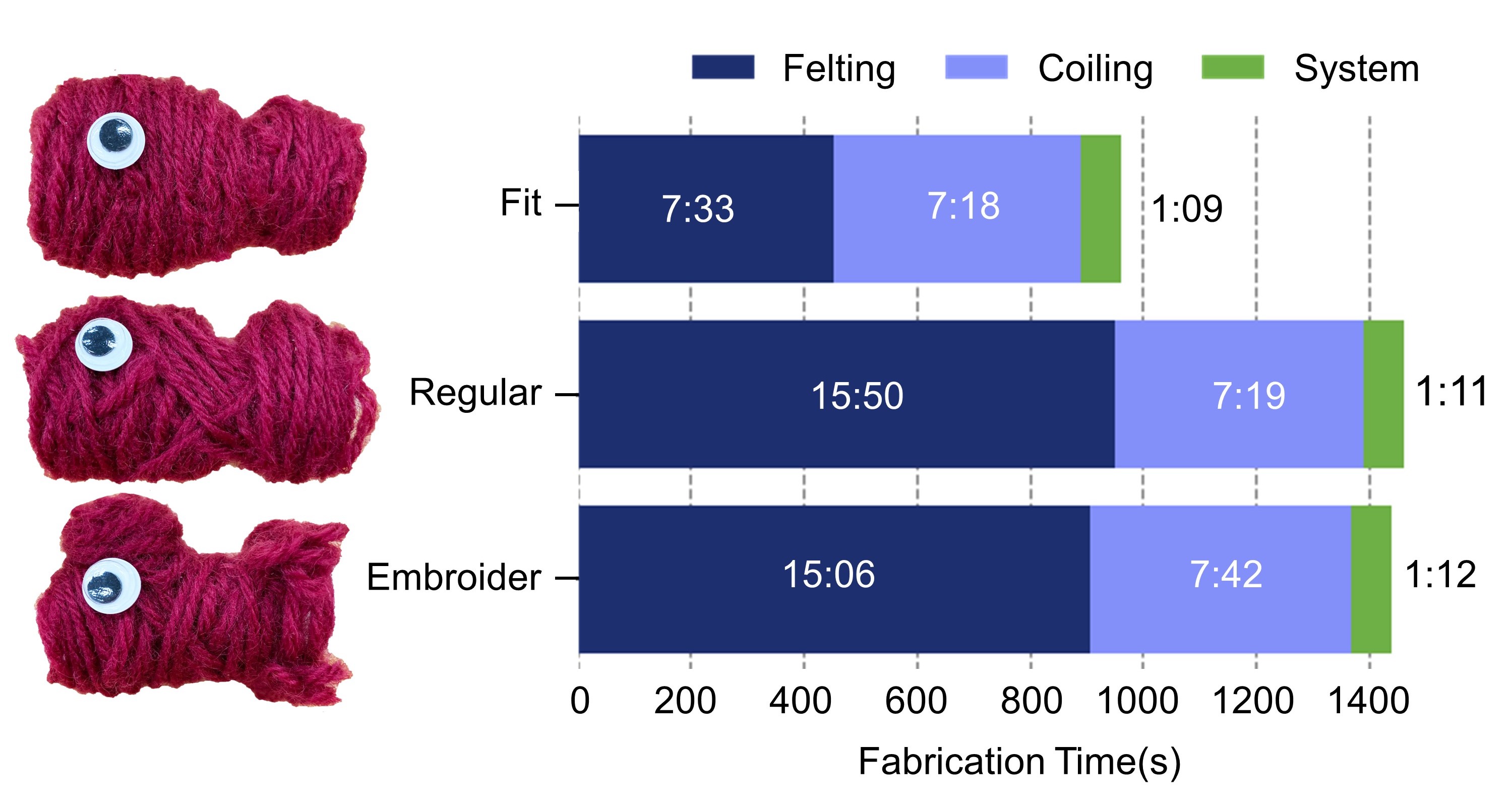}
    \caption{ We compared the fabrication times of fish dolls made using three different methods. The coiling time for all three cases is similar, while the fit coiling method takes less time for felting. }
  \Description{Figure 7 shows the bar plot of the fabrication times of a fish doll made using three different methods: fit coiling, regular coiling, and embroidered. The right of the bar demonstrates the three workpieces made by these methods. The felting time is 7:33, 15:50, and 15:06, respectively; the coiling time is 7:18, 7:19, and 7:42; the system time is 1:09,1:11 and 1:12.  }
  \label{fig: fish visualization}
\end{figure}

\autoref{fig: phone visualization} and \autoref{fig: fish visualization} show the phone and fish shape examples' fabrication times and results using three methods. 
Regular coiling is the basic method to form the workpiece, but it takes more felting time.
While the regular cylinder formation is faster than the fit coiling method, the system may require more time to compress the model if it contains deep concave shapes.
The regular coiling method takes more time to fabricate the whole workpiece in these two examples. 
In general, the regular coiling method is suitable for models having gentle curves.

Fit Coiling reduces the fabrication time, especially in the felting time in these two cases. 
The coiling base shape before felting influences the felting time a lot. 
The system can shorten the felting time more if the target model's shape is nearly the rotational symmetric model.

The embroidered method is suitable for models with protruding parts, especially those parts that are not rotationally symmetric. 
This method spends less time coiling, while the result is much rougher than the other two methods. 
Using felting as a refined method requires a considerable amount of time. 
Additionally, to achieve the same level of smoothness as other methods, a significantly longer felting time is necessary. 
It is suitable for rapidly forming the prototype of the workpiece.

\section{Evaluation}
This section primarily focuses on the factors that impact the density of the soft material, specifically the influence of coiling tension, coiling techniques, felting times and needle tips. 
Additionally, we present our soft material selection process.

\subsection{Material Selection}
Before fabricating the soft material, we search for a suitable material, which is easier to compress and can be tangled together to use in our system, so we conduct a material test. 
The thread can be generally categorized into three by material source: animal, plant, and artificial (chemical) fiber. 
In this test, we use four kinds of thread to represent the different sources: Wool Yarn (animal fur), Hemp Rope (plant fibers), Acrylic Line (artificial fibers), and Ribbon (artificial woven fibers).
In the test, we coil a similar quantity of the different fibers to a ball and manually felt each of them 200 times to observe the volume variation of the thread. 

\begin{figure}[htb]
\includegraphics[width=1\columnwidth]{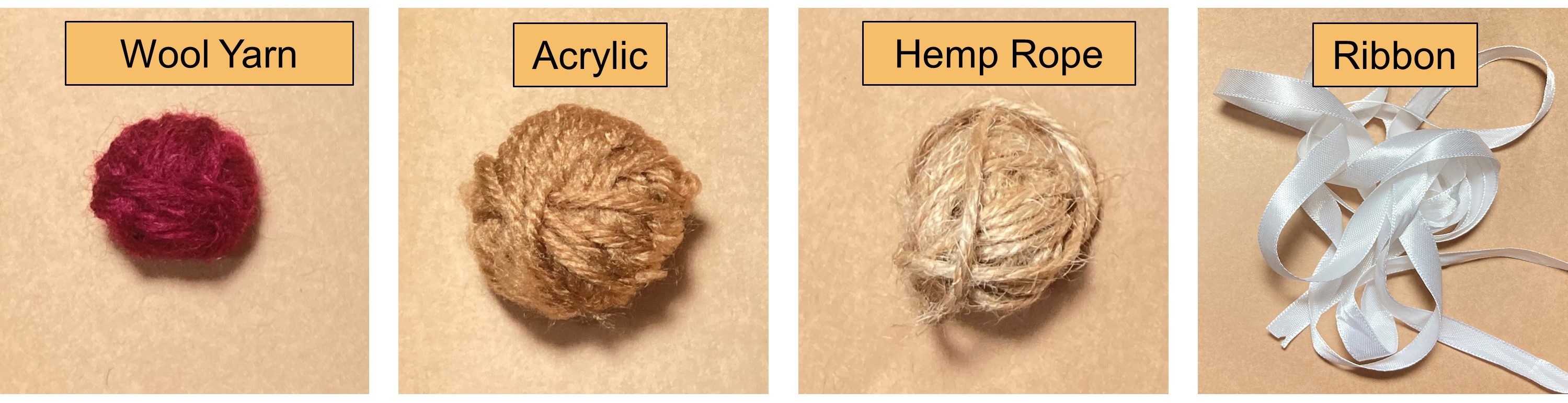}
    \caption{ The results of the material test by coiling to a ball and then felt it. The wool yarn has the best performance at the compressed and entangled levels. Acrylic and hemp rope' performances are not well enough in compression, and ribbons can not entangle together.}
  \Description{ Figure 8 shows the felting results of the material selection test. From left to right (wool yarn, acrylic line, hemp rope, and ribbon) shows the compressed and entangled level from high to low. The wool yarn has the best performance at the compressed and entangled levels. Acrylic and hemp rope' performances are not well enough, and ribbons can not entangle together.  }
  \label{fig: material}
\end{figure}

The results in \autoref{fig: material} show the corresponding entanglement and compressive levels.  
Due to tiny splits on the surface of the yarn, these micro-structures allow for easier entanglement of fibers compared to other materials tested.
Hemp rope and acrylic can also be tangled together, but the connection between the fibers is weaker than wool yarn, it needs more stabbing times to fixate the fiber. 
The acrylic line can also be compressed after felting but can not compress as well as the wool yarn. 
Hemp rope, which is more tightening than others, does not contain the space to be compressed. 
The fibers of the ribbon cannot become entangled during testing because the thread is arranged too uniformly and lacks the flexibility to interact effectively with the fibers underneath it.
In addition, the loose structure of yarn allows it to be compressed the most tightly among these four materials.
According to the test of entanglement and compression's degree, we choose yarn as our primary material to use in the system. 
It will take less time to fixate the fiber and shape the object.

\begin{figure}[htb]
  \includegraphics[width=1\columnwidth]{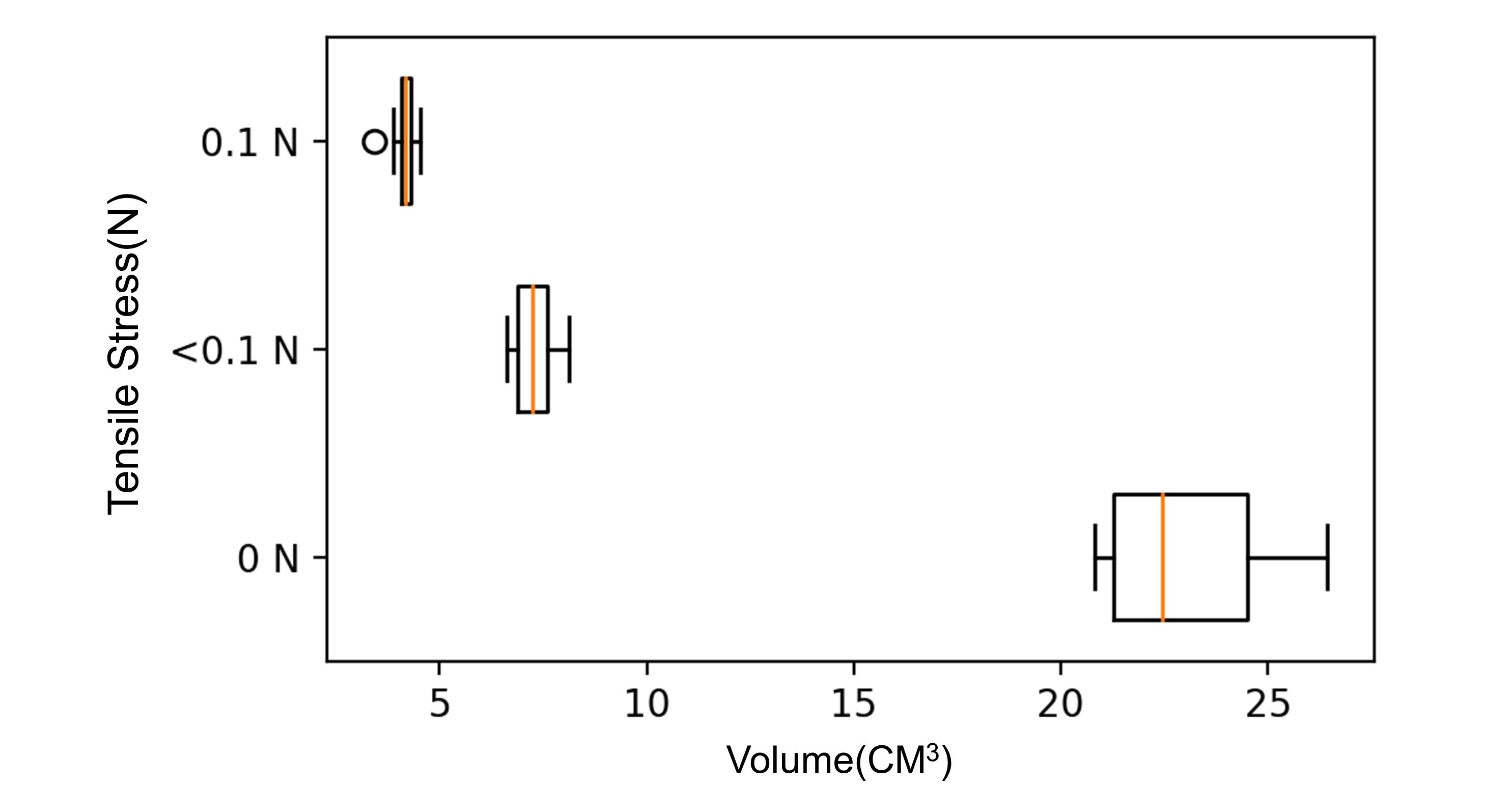}
    \caption{ The boxplot displays the volumes of five examples for each coiling tension level. The chart shows that the less tensile exists, the larger the range of volume is. }
  \Description{ Figure 9 shows the boxplot that displays the volumes of five examples for each coiling tension level. The tensile stress is 0.1N, <0.1N, and 0N. The chart shows that the less tensile, the larger the volume range is.}
  \label{fig: tension volume}
\end{figure}

\subsection{Coiling Tensile Test}
In this test, our goal is to know how the tensile stress of the line will influence the results of the coiling structure. 
To control the coiling tensile stress, we change the number of moving steps of the two stepper motors just like changing the gear ratio in our system.
In the test, we coil cylinders with three different tensile, each tensile with 5 repeating tests). 
The tensile stress is none (0N, loose), low (smaller than 0.1N, straight), and high (about 0.1N, which is tight enough for our system to compactly coil the line onto the reel.) 
The force is detected by a tensiometer.
Although the force is only smaller than 0.1N, the force is enough to have a different result. 
If the force is larger than 0.1N, the thread will coil on the reel tightly, which cannot be shaped by coiling and felting.
The results in \autoref{fig: tension volume} show that the less tensile exists, the larger the range of volume is.
For no stress, it is less controllable due to the low friction, and the density is lower than the other two ratios. 
The low tensile stress is tight enough to control coiling, and its result is flexible to be tuned. 
Due to the result of the three stress, we use low tensile stress in our system to provide a stable coiling structure and keep the room to be tuned by the felting machine.  

\begin{figure}[htb]
  \includegraphics[width=1\columnwidth]{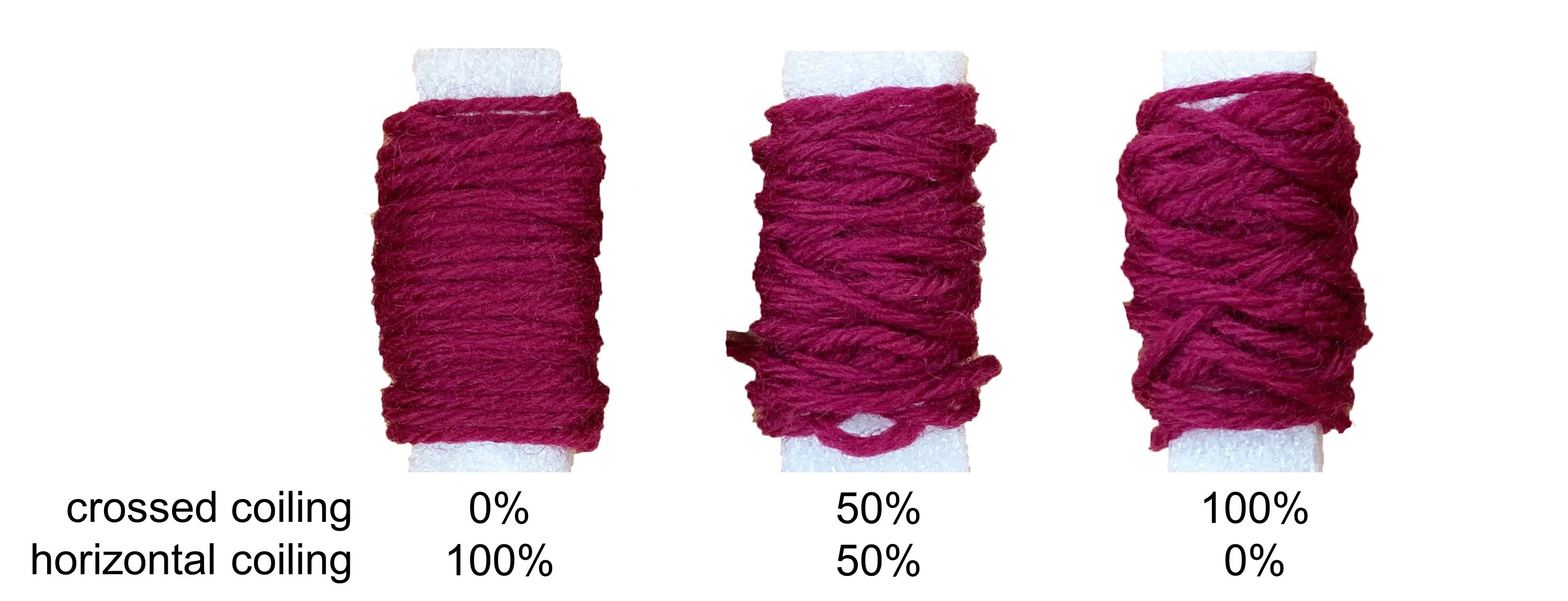}
    \caption{ The figure demonstrates the cylinder results made with different crossed coiling ratios in the coiling way test.}
  \Description{ }
  \label{fig: coiling way}
\end{figure}

\begin{figure}[htb]
  \includegraphics[width=1\columnwidth]{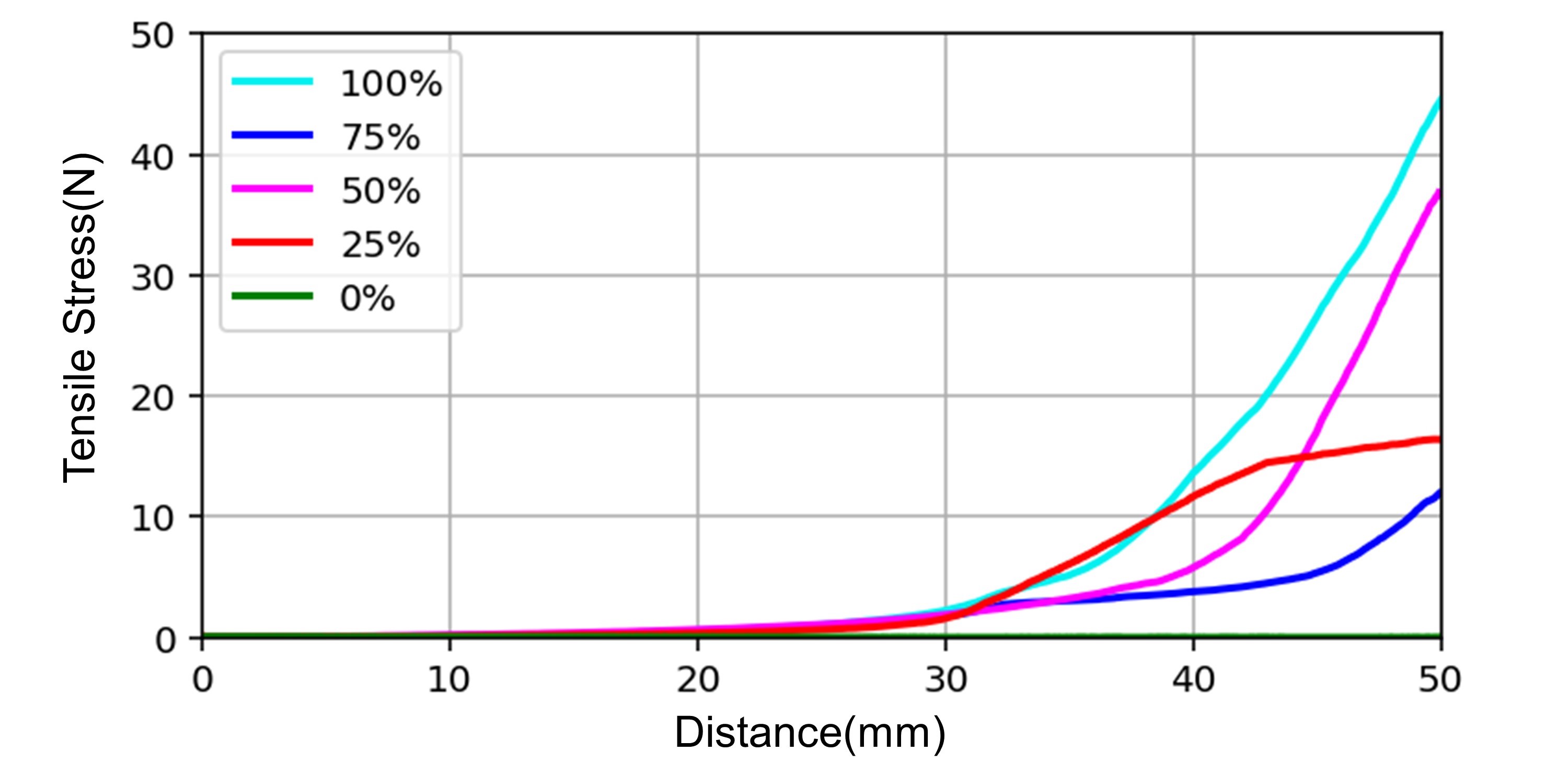}
    \caption{The chart demonstrates the tensile stress variation of different crossed coiling ratios through the deformation distance. The results using 100\% and 50\% crossed coiling have better performance in enhancing structural strength. }
  \Description{ Figure 10 shows the line chart of the tensile stress variation of different crossed coiling ratios (100, 75, 50, 25, 0) through the deformation distance (50mm). The results using 100\% and 50\% crossed coiling have better performance in enhancing structural strength. }
  \label{fig: crossed ratio tensile}
\end{figure}

\begin{figure}[htb]
  \includegraphics[width=1\columnwidth]{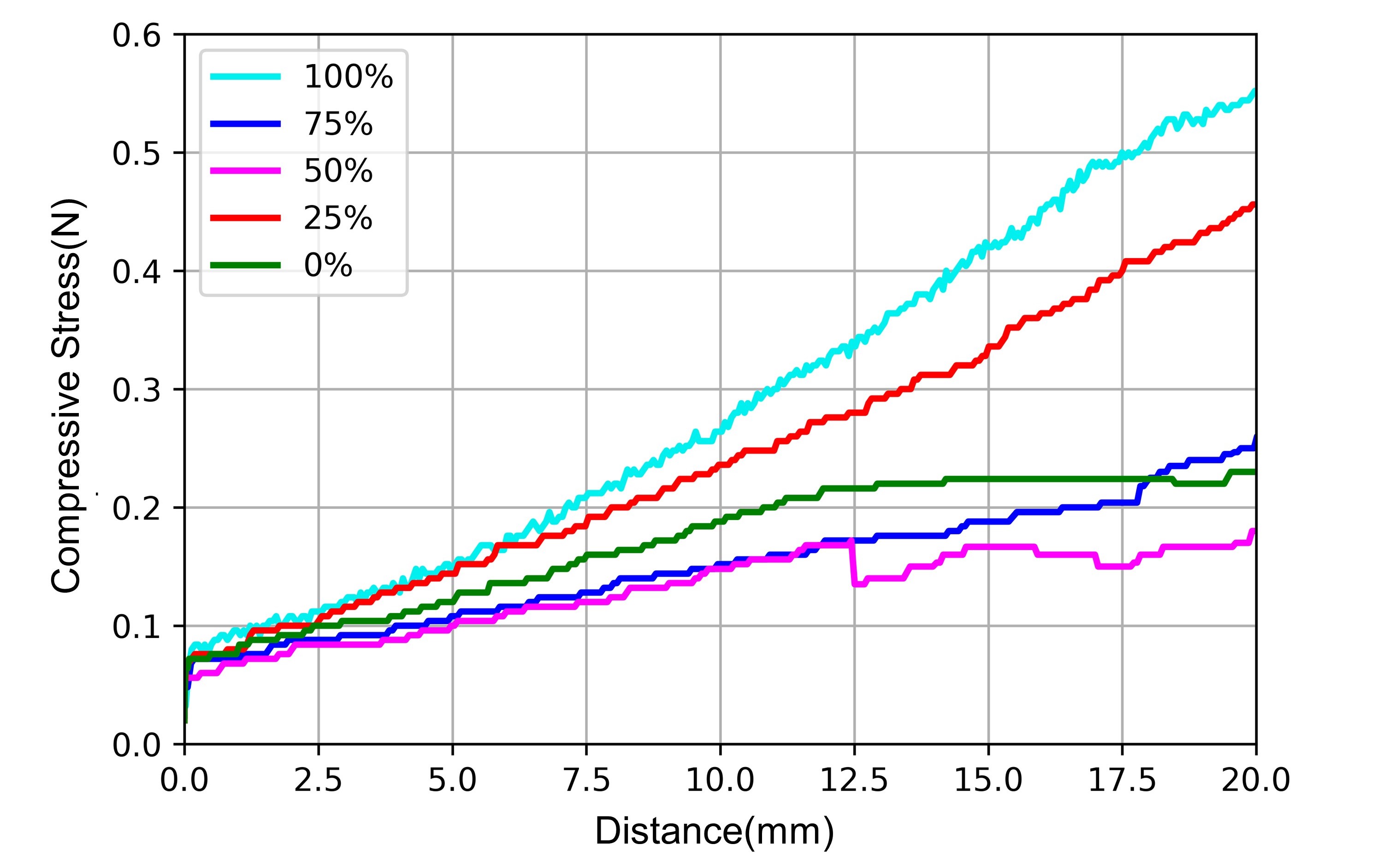}
    \caption{The chart demonstrates the compressive stress variation of different crossed coiling ratios through the deformation distance. The results using 100\% crossed coiling have the most compressive stress variation. }
  \Description{ Figure 11 shows the line chart of the compressive stress variation of different crossed coiling ratios through the deformation distance. The results using 100\% crossed coiling have the most compressive stress variation. }
  \label{fig: crossed ratio compressive}
\end{figure}

\subsection{Coiling Way Test}
Besides the influence of tensile during the coiling process, we also want to understand how the coiling way influences the structure of results. 
Our system supports two coiling ways: horizontal coiling and crossed coiling. 
The horizontal coiling coils the line horizontally to the reel like a spring. 
The crossed coiling pattern is formed through the moving of the Stewart platform. 
The platform moves up and down to control the slope of the line between the core and the feeder.
In this test, our goal is to evaluate the structural strength influenced by these two coiling ways.

During the test, we use the same number of cycles to see the effects of mixing these two coiling ways.
We test 5 different crossed-coiling ratios, and each work's total cycles are 200 cycles. 
0\% refers to 200 cycles of horizontal coiling. 
25\% ratio means 5 cycles of crossed coiling plus 15 cycles of horizontal coiling and then repeating the pattern 10 times. 
50\% ratio means 10 cycles of crossed coiling plus 10 cycles of horizontal coiling and then repeating the pattern 10 times. 
The remaining ratios are 75\% and 100\%, and the relation cycles follow the rule above. 
The maximum slope of the line in the system is 60 degrees. 

We use the stress-strain measurement machine to test each coiling structure's tensile stress and compressive stress. 
Tensile stress demonstrates the capacity to enhance structural strength, and compressive stress reflects the flexibility and softness of workpiece.
\autoref{fig: crossed ratio tensile} shows the tensile stress variation of each different crossed coiling ratio structure through the stretching distance variation up to 50 mm. 
If the stretching distance is too long, the coiling structure will unfasten to a loosened yarn, and the tensile stress measured by the machine will not be the coiling structure but only the yarn material itself. 
The structure of 100\% crossed coiling has the strongest structural strength than other structures, which can get up to 44.6N when stretching the coiling object to 50 mm. 
Half-crossed coiling and half-horizontal coiling combination also perform well in tensile stress testing. 
The 25\% and 75\% crossed coiling ratio results' structural strengths are half lower than those mentioned above. 
If only coil the line horizontally without felting the object, the structure is unstable and easy to loosen to a string. 
As a result, the stress measurement in tensile stress testing is lower than 1N, even if stretching to 50 mm. 

For the compressive stress test, we use the machine to bend the samples to 90 degrees which is the distance of moving 20 mm to the machine to get the pressure data. 
The data in \autoref{fig: crossed ratio compressive} shows that all samples' compressive stresses are less than 1 N. 
The reason is that the testing samples are made from yarn, which is a soft material, and its strength is not enhanced by the felting. 
Due to the property of the soft material, the fiber has the flexibility to be bent. 
Hence coiling ways do not gain too much resistance to compressive forces.
As a result, if we want to have a strong structural strength of a certain part of the workpiece, the part should be coiling with a higher crossed coiling structure.
If we want to have a bendable structure like the elbow part of the sleeve, we can use a lower crossed coiling structure. 
Depending on the function or desired result, the system can coil different structural strengths by different ratios between crossed coiling and horizontal coiling. 

\subsection{Felting Needle Tips Test}
\begin{figure}[htb]
  \includegraphics[width=1\columnwidth]{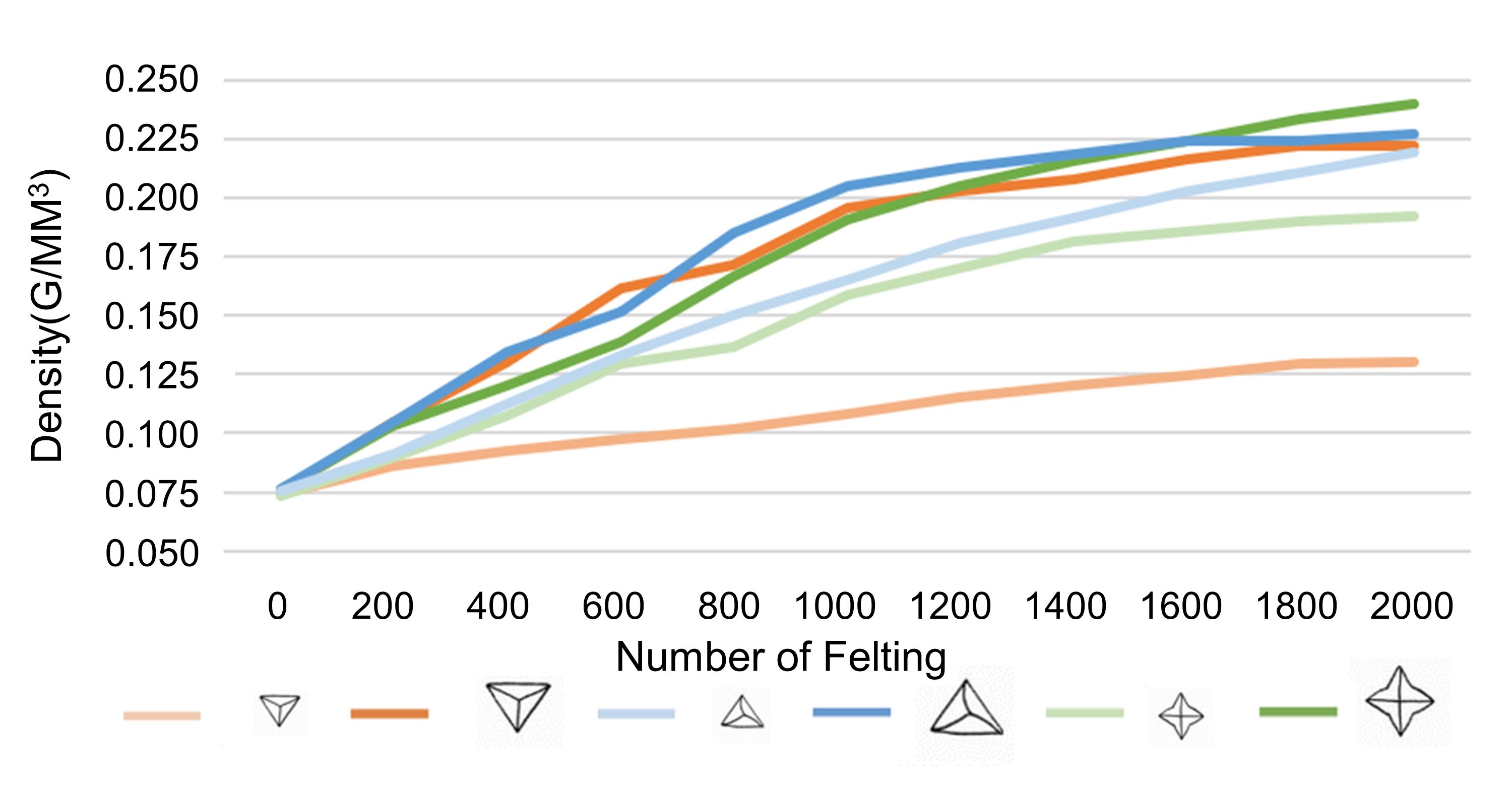}
    \caption{ The wool yarn density variation with three different felting needle tips plus two thicknesses. The three icons in the image indicate the tip shapes of the felting needles, which are triangle, spiral, and star from left to right. Each shape has two sizes, with the larger size indicating a thick needle and the smaller size representing a thin needle.}
  \Description{ Figure 12 shows the wool yarn density variation line chart with three different felting needle tips plus two thicknesses. The three icons in the image indicate the tip shapes of the felting needles, which are triangle, spiral, and star from left to right. Each shape has two sizes, with the larger size indicating a thick needle and the smaller size representing a thin needle. The density is from 0.075 to 0.225g/mm^3. }
  \label{fig: needle density}
\end{figure}

There are different shapes of the felting needle tip and thickness of the needles. 
In this test, our goal is to test how the different needle tips will influence the density and structural strength of the punching results.
We choose three common felting needle tip shapes (Triangle, Star, Spiral) with two thicknesses of each needle for testing the influence of the density and strength variation. 
We use the three icons in the image to indicate the tip shapes of the felting needles, which are triangle, star, and spiral from left to right in \autoref{fig: needle density}.
Each shape has two sizes, with the larger size indicating a thick needle and the smaller size representing a thin needle. 
We use these icons as the legends of charts. 

In the test, we put 1g wool yarn into a cuboid with a 2.5*2.5cm bottom area and felt it equally in this small area with different needles. 
We use a caliper to measure the altitude change after felting multiple times. 
We calculate the density variation by knowing the yarn's mass and the volume's variation. 

\begin{figure}[htb]

  \includegraphics[width=1\columnwidth]{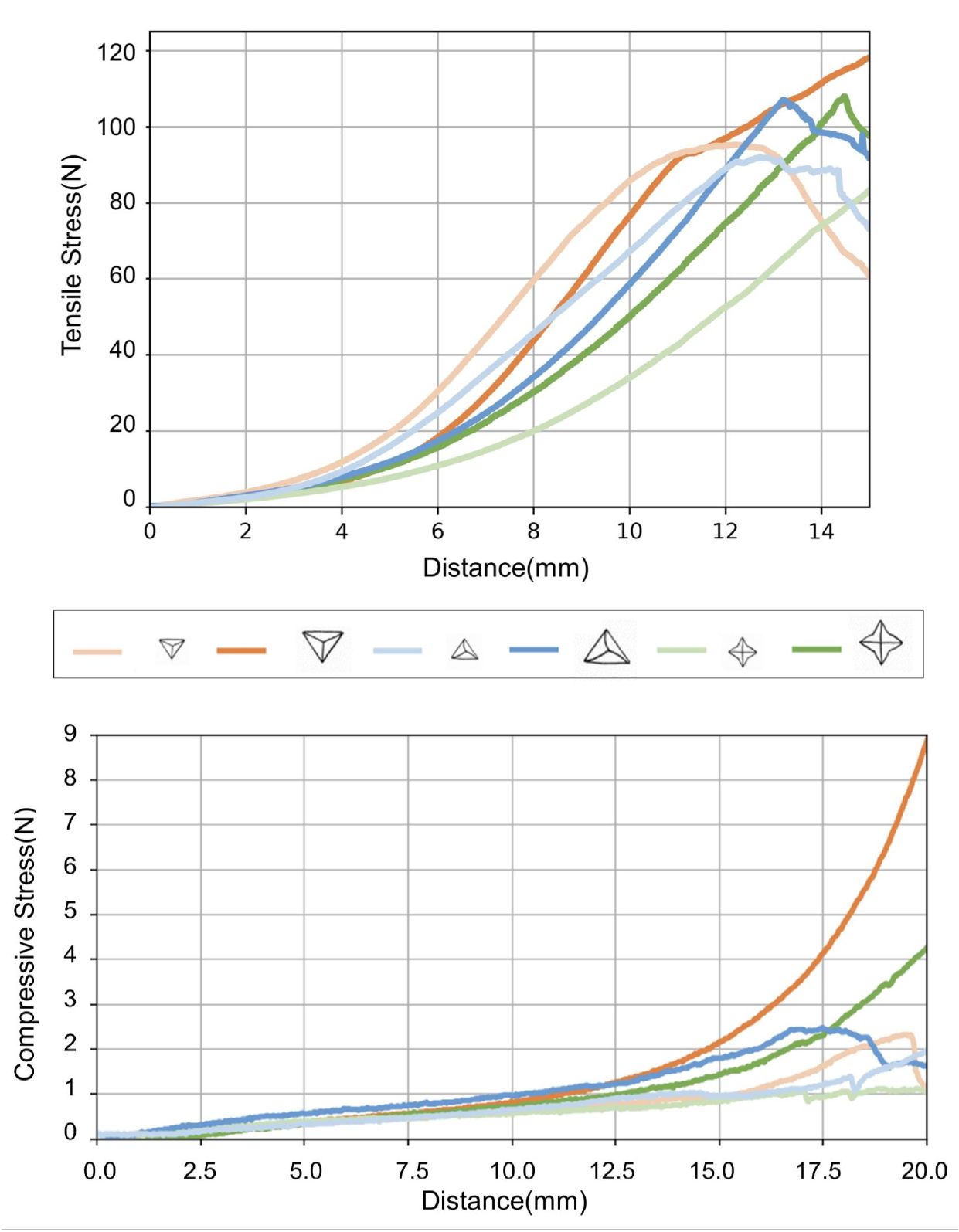}
    \caption{ UP: The tensile stress variation of cuboids made from different felting needles tips. Down: The compressive stress variation of cuboids made from different felting needles tips. The cuboids' tensile stress is larger than the compressive stress. The compressive stress of all the cuboids increases slowly during the first half deformation distance. }
  \Description{ Figure 13 shows the tensile stress variation of cuboids made from different felting needles tips (up) and the compressive stress variation of cuboids made from different felting needles tips (down). The cuboids' tensile stress is larger than the compressive stress. The compressive stress of all the cuboids increases slowly during the first half deformation distance. }
  \label{fig: needle stress variation}
\end{figure}

\begin{figure}[htb]
  \includegraphics[width=1\columnwidth]{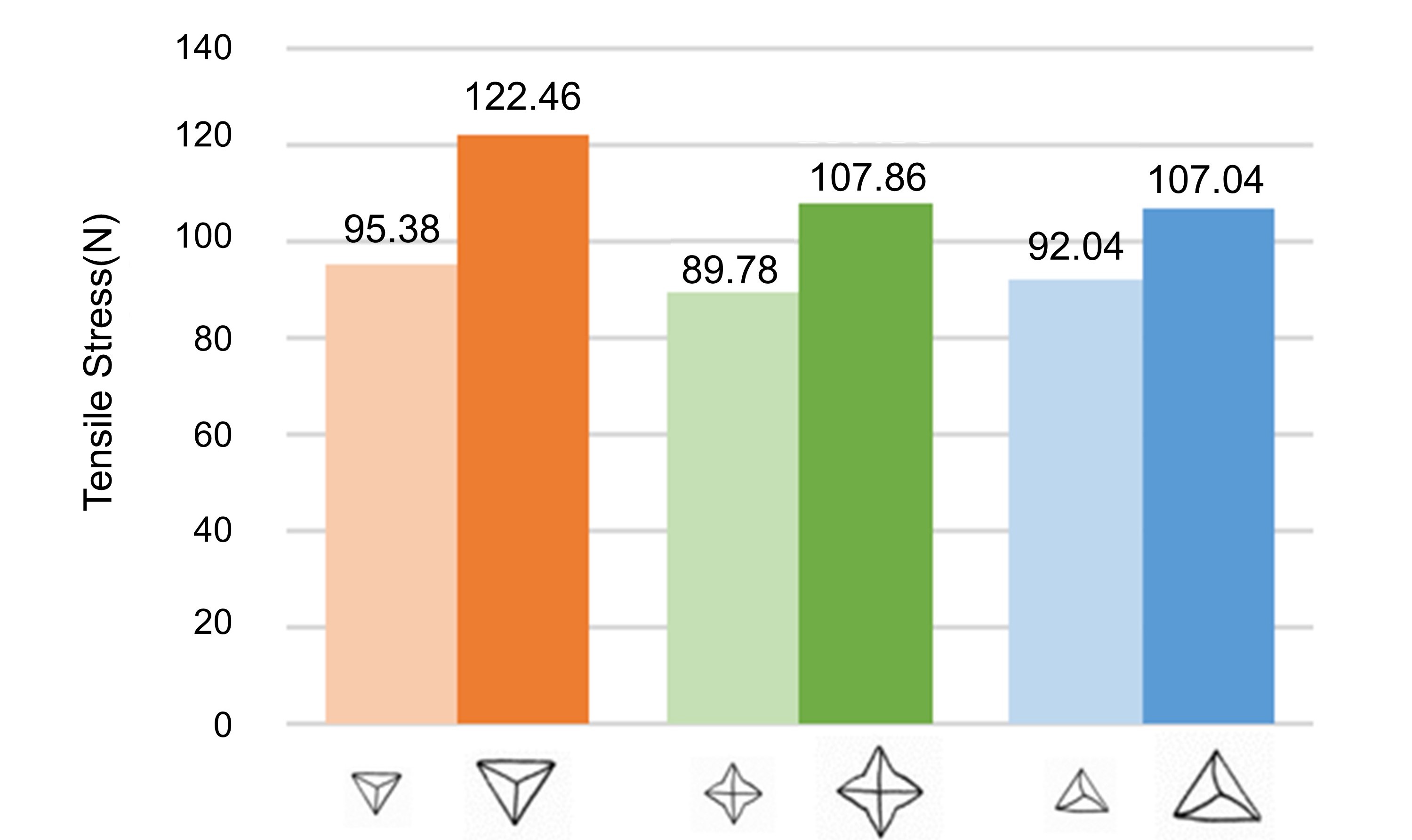}
    \caption{ The max tensile stress we get from the cuboids made by different felting needles tips during the deformation. Thicker needles have better performance. }
  \Description{ Figure 14 shows the max tensile stress we get from the cuboids made by different felting needles tips during the deformation. Below are the tensile stress: 95.38N (Triangle Thin), 122.46N (Triangle Thick), 89.78N (Star Thin), 107.86N (Star Thick) , 92.04N (Spiral Thin) , 107.04N (Spiral Thick).}
  \label{fig: needle tensile}
\end{figure}

\begin{figure}[htb]
  \includegraphics[width=1\columnwidth]{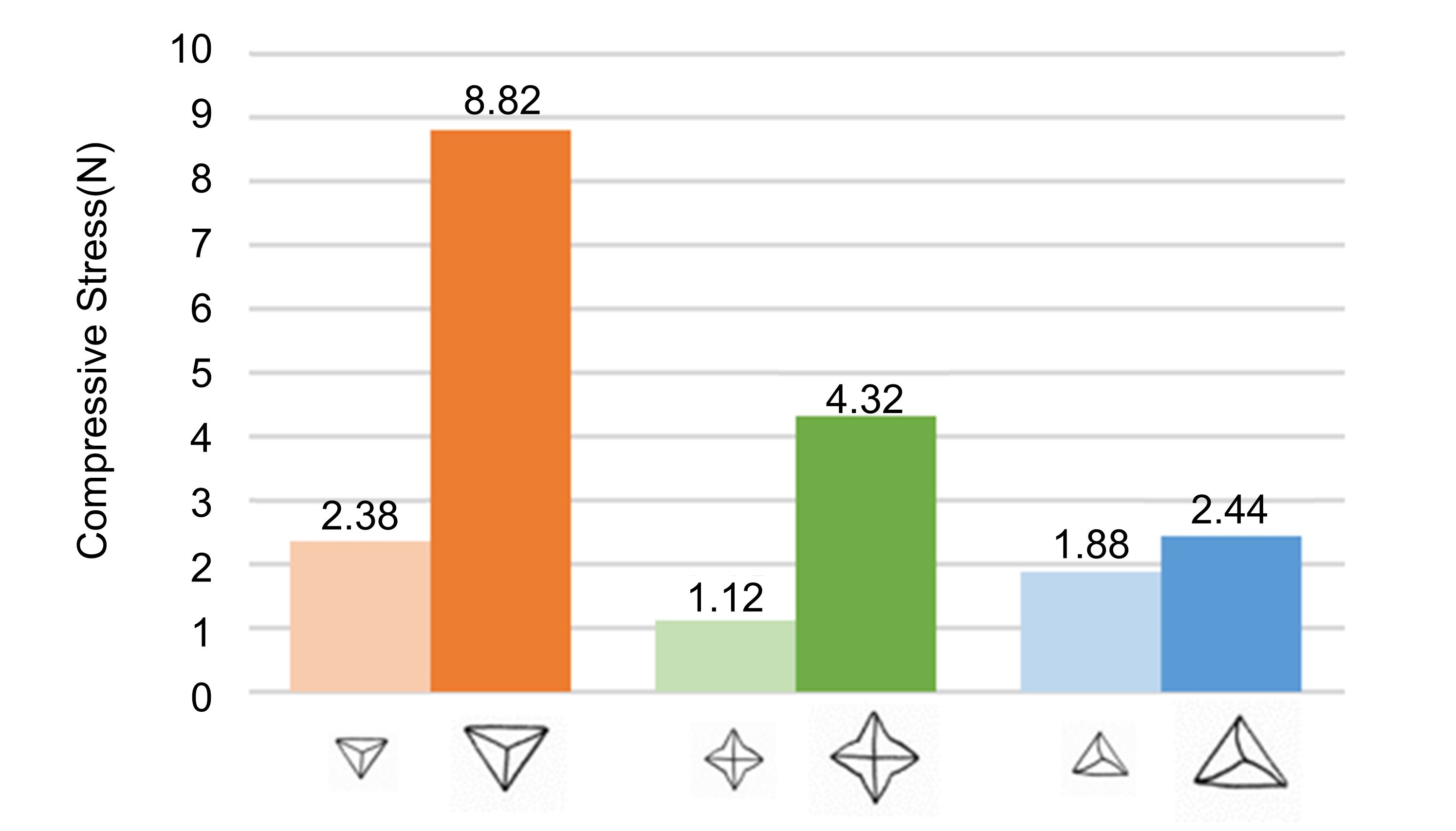}
    \caption{ The max compressive stress we get from the cuboids made by different felting needles tips during the deformation. The thick needle with the triangle shape has the best performance. }
  \Description{ Figure 15 shows the max compressive stress we get from the cuboids made by different felting needles tips during the deformation. Below is the compressive stress: 2.38N (Triangle Thin), 8.82N (Triangle Thick), 1.12N (Star Thin), 4.32N (Star Thick), 1.88N (Spiral Thin), 2.44N (Spiral Thick).  }
  \label{fig: needle compressive}
\end{figure}

The result in \autoref{fig: needle density} shows that all of the three thickness needles have quicker density variation and higher density increment. 
The thickness of the needles influences more than the shapes of the needle tips. 
Comparing the performance between the shape of the needles, the triangle needle tips compress the fiber faster before 800 times felting. 

Besides recording the density variation, we also test the tensile and compressive stress to show how felting enhances structural strength. 
We first test the compressive stress to avoid the samples' deformation during the tensile stress test. 
In the compressive test, we use the stress-strain machine to bend the samples, as same as the testing process in the coiling method strength test. 
The distance of testing is 20 mm. 
The bottom of \autoref{fig: needle stress variation} demonstrates the variation of the compressive stress. 
The material property lets the stress increase slowly to 1N, and the curve is gentle until bending to 10 mm. 
After bending to 10 mm, the structural strength starts to reflect on the compressive stress, and the stress starts to grow violently. 
The samples made from thick needles have higher density and gain more structural stress after 10 mm than the thin needles sample. 

The chart at the top of \autoref{fig: needle stress variation} displays the variation of tensile stress. 
The curves, which represent the thick star-shape and thick spiral-shape needle, arrive at the maximum tensile stress at 13.2 mm and 14.25 mm and then begin to decrease. 
Unlike these thick needle samples that decrease more rapidly to form a peak in the curve, the turning of the curves represents the thin needles forming a more smooth peak. 
We consider this difference happens because the thick needle at the felting process forms a stronger connection between the fiber than using a thin needle, so it can achieve higher stress. 
Still, when achieving maximum stress, the connection is broken quickly. 
The surrounding connection is also broken quickly. 
While the fiber connection formed by the thin needle is not stronger, it is more even. 
That is, when it achieves the maximum tensile stress, the force can be scattered to the surrounding entanglement point, forming a smoother peak. 

\autoref{fig: needle tensile} and \autoref{fig: needle compressive} demonstrate the maximum stress value in the tensile and compressive stress test. 
In the charts, we clearly see that the thickness of the needles influences more than the shape of the needle tips. 
By comparing the performances in these tests, we select the thicker triangle needle tip as the needle shapes used in our system.
By knowing the density variation, we can know how many times we should felt to the workpiece to control the density variation.

\section{Results}
\begin{figure}[htb]
  \includegraphics[width=1\columnwidth]{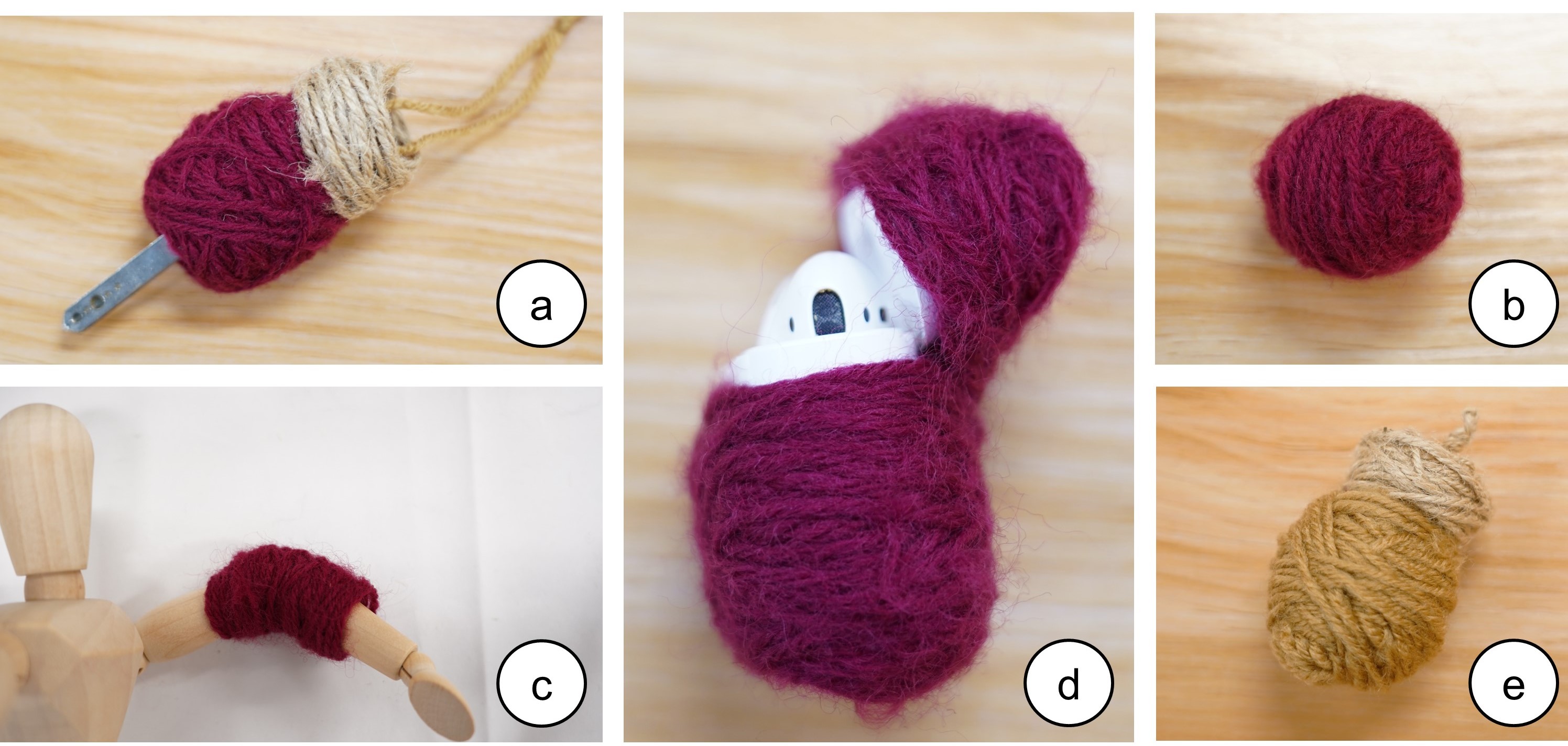}
    \caption{ We demonstrate a combination example of yarn and hemp rope as (a) a key case and (e) an acorn, and other results: (b) a sphere (c) a bendable sleeve for a doll (d) the side view of an earphone's case.}
  \Description{ Figure 16 demonstrates the combination examples of yarn and hemp rope as (a) a key case and (e) an acorn, and other results: (b) a sphere (c) a bendable sleeve for a doll (d) the side view of an earphone's case. }
  \label{fig: results}
\end{figure}

We demonstrate the capability of our system by using it to fabricate various 3D models with changing reel core shapes and using different materials.
The results are shown in \autoref{fig: results}.
The main fabrication procedure has been described in Section 1, so we only highlight the specific steps used for each result in this section.

The cylinder in \autoref{fig: stream}c is our system's basic felting base when making objects. 
It takes about 3-10 minutes to form the result, depending on the model's size, desiring tightening level, and the width of the material. 
We can use it as a customized doll's sleeve (\autoref{fig: results}c) or pencil sheath after we remove the reel core to form the hollow cylinder. 
We can loosen the structure of the bending part to provide more room for bending. 

The sphere shown in \autoref{fig: results}b is formed by coiling the line from different angles while keeping the intersected center at the sphere's core.
We then control the position and tilt of the felting machine to refine its arc. 
In our system, the majority of the time is spent on refining the work, and it can take approximately 5-15 minutes to achieve the desired arc.
To reduce the required time, we first coil the thread to form a shape close to the desired model as we described in \autoref{sec:3D}.

We use the earphone case size as the reel core when fabricating the earphone case cover shown in \autoref{fig: teaser}c and \autoref{fig: results}d. 
When forming the flippable top lid, the system controls the motion platform tilt up and down to make the line cross on the same side. 
We coil higher on the bottom side for the short cap and the larger body. 
We strengthen the crossing side (\autoref{fig: results}d) to prevent the bending part from scattering. 
After removing the workpiece from the system, we manually use the needles to close the top lid and decorate it with surplus yarn when coiling the top or the additional yarn.
The system takes less than 10 minutes to form the case cover, and we spend about 2 minutes post-processing the lid.

Changing the thread spool can also form multiple colors or material works.
We build the connection between the two materials by felting the overlapping part. 
After coiling, the connection part disappears in the entangled fibers. 
The system can be used with different materials for various functions. 
For example, yarn is a good material for bending and is easy to tangle together, and hemp rope is firmer and suitable for strengthening the bottom if we want the workpiece to stand stably. 
Using different textures can add more variation to the workpiece.
For example, \autoref{fig: results}a use the hemp rope to strengthen the top of a key case, and \autoref{fig: results}e uses the hemp to form the lid of the ellipsoid and then manually adds extruded hemp to make it like an acorn.
These different materials can be used in the workpiece to change certain parts textile for enhancing the structure or decoration. 
The use of different materials requires more time to fixate or shape.

\begin{figure}[htb]
\includegraphics[width=1\columnwidth]{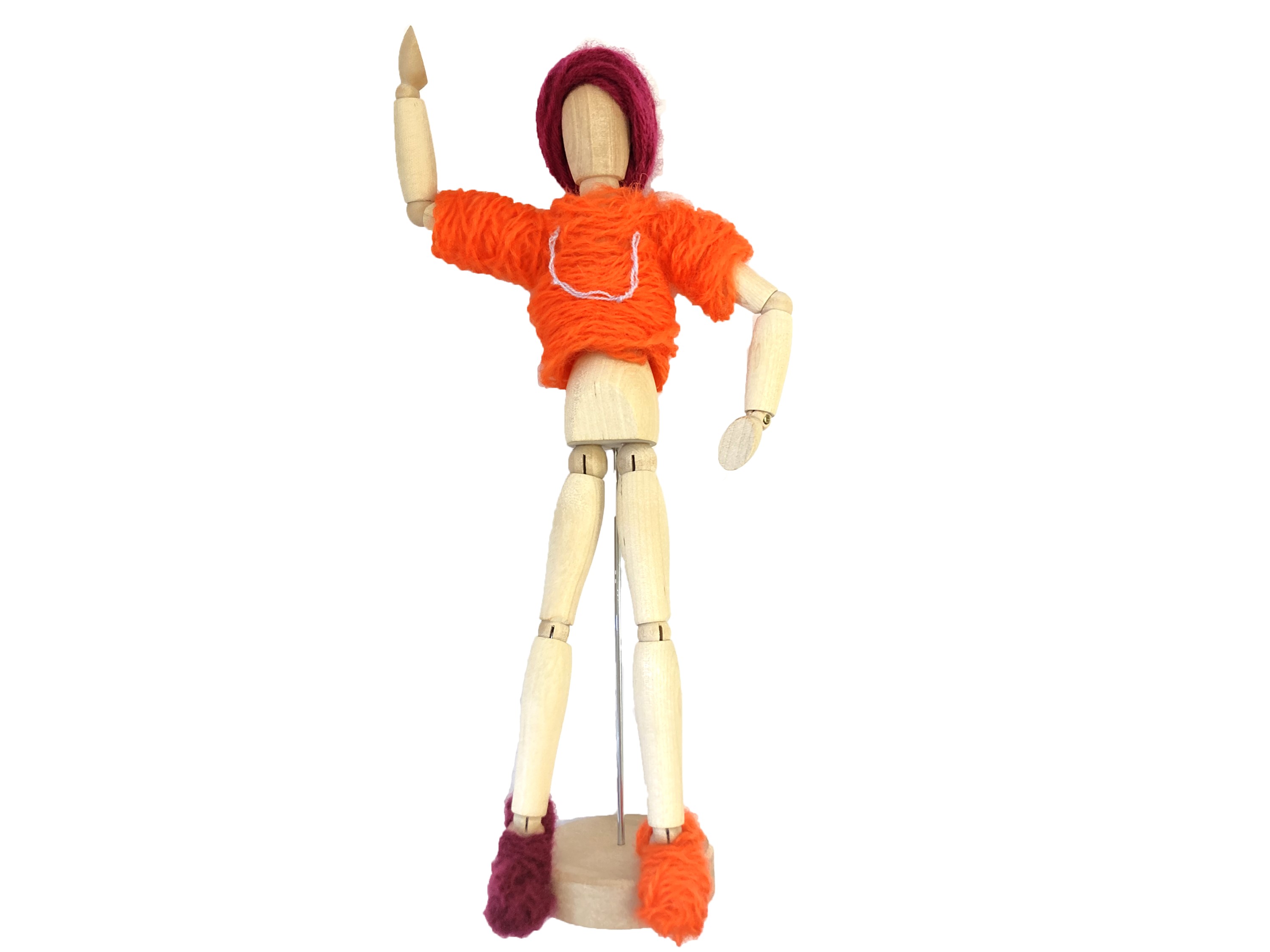}
    \caption{ We change the shape of the cores to let the system make the hat, T-shirt, and shoes fit the doll's head, body, and feet. We combine the sleeves to the vest to form a T-shirt by felting the connection parts. We also decorate the T-shirt by felting the line on it. We use the puppet and the wearing on itself to show the possibility of quickly making the prototype of the textile. }
  \Description{ Figure 17 shows a puppet wearing a red hat, an orange T-shirt with a thinner line embroidered on it forming a letter 'U', a red shoe on the right foot, and an orange shoe on the left. }
  \label{fig: doll}
\end{figure}

We provide more variations of shapes made by the system in \autoref{fig: doll} to showcase its capabilities.
To make the hat, T-shirt, and shoes fit the doll's head, body, and feet, we change the shape of the cores.
We also combine the sleeves to the vest to form a T-shirt by felting the connection parts, and we decorate the T-shirt by felting a line on it.
The T-shirt demonstrates the system's ability to bridge between each component, allowing for larger works to be created by bridging multiple components together.
In addition, the shoes demonstrate using different line widths and colors to form the same model.
The bottom parts of the shoes have been felted more times to form solid soles while keeping the top side of the shoes more flexible.
This is achieved using a looser coiling structure and less felting, allowing easier wear onto the puppet.
We have used the puppet and shown how it can be used to create a textile prototype quickly.

All these works showcase the diversification of our system. 
We demonstrate the basic models, the hollow object, the workpiece with different core shapes, the preprocessing work, and the combination of the workpieces.

\section{Discussion, Limitation, and Future Work}
While coiling can be used to form the 3D shape, it requires additional fixative methods to prevent the thread from scattering.
The main drawback of coiling is the limited ability to control the yarn's placement since only friction is used to position it.
The friction between the coiled material and the surface it rests on can only provide slight stability.
As a result, more fixative techniques are typically required to achieve precise placement and maintain the shape.
However, felting can be used to strengthen and stabilize the coiled structure, compensating for this limitation.

The wool roving (non-spun fibers that are typically provided in a band or sheet and used in felting craft) is easier to shape, while it is not easy to feed quantitatively in the device. 
One possible way to use wool roving would be a technique like spun-bond (melt the chemical material and then spray them to a base to form the mesh of fiber) to control the quantity of the fabric and fixate the fiber together when cooling down. 
An alternative approach would be to use a machine that works similarly to a cotton candy maker to spray the wool and create a printed object on a rotating axle.

The system can be a mixed-initiative or interactive fabrication system.  
After creating the basic model, users can choose whether to revise the shape or coil more material onto the base. 
If users want to refine the model, users can control the motion platform to let the felting machine stab on the desired part and decide the times of felting to adjust the compressed level of fiber. 
If users want to coil more material, users can feed more yarn to the desired part. 
Users can keep revising the shape or coiling more material until they are satisfied with the result. 

We can improve the system by adding more digital assistance. 
For example, the setup steps can be automated with improved mechanical design. 
A user interface for interactive fabrications is also on our bucket list. 
Currently, we manipulate the platform and felting machine directly in the refining process. 
The system can recommend to the user how to revise the model without learning what angle to stab or how long it takes for felting to form the desired shape. 
Furthermore, we can add sensors or cameras to the device to detect the work's shape in real-time to achieve a close-loop fabrication system.

\section{Conclusion}
In this work, we introduce \projectName{}, a soft fabrication system that combines coiling and felting techniques to create objects with various structural strengths.
We evaluate the effects of different felting needles, frequencies, and coiling patterns that influence the density, structural strength, and fabrication time of the workpiece.
Through our experiments, we demonstrate \projectName{}'s ability to change the density of the object, providing more freedom for 3D soft fabrications.
This enables us to strengthen specific parts of the object and optimize its functionality.

\begin{acks}
This work was supported by the National Science and Technology Council in Taiwan (112-2636-E-002-002). 

To my esteemed mentor, I am immensely grateful for your invaluable guidance and generous investment of time throughout this project. Your expert advice has been instrumental in shaping this endeavor into something truly remarkable.

I extend my heartfelt appreciation to all the members of the HCI lab. Our collaborative discussions have played a crucial role in bringing this project to fruition, and I am thankful for the collective effort and support that has been poured into it.

A special thank you goes out to Ming-Hang for the indispensable assistance provided in building the hardware.

Last but not least, I am deeply thankful to everyone who believed in and supported this project. Your encouragement and enthusiasm have been a constant source of motivation.
\end{acks}

\bibliographystyle{ACM-Reference-Format}
\bibliography{reference}


\end{document}